\documentclass[submission,Phys]{SciPost}

\usepackage{relsize}
\usepackage{physics}
\usepackage{bbold}

\usepackage[utf8]{inputenc} 
\usepackage[T1]{fontenc} 	
\usepackage[english]{babel} 

\usepackage[bitstream-charter]{mathdesign}
\usepackage{geometry} 		
\usepackage{amsmath} 		
\usepackage{mathtools} 		
\usepackage{float} 			
\usepackage{graphicx} 		
\usepackage{tabularx} 		
\usepackage{booktabs} 		
\usepackage{color, xcolor} 	
\usepackage{pdfpages} 		
\usepackage{extarrows} 		
\usepackage{multirow} 		
\usepackage{multicol} 		
\usepackage{subcaption} 	
\usepackage{enumitem} 		
\usepackage{xspace} 		
\usepackage{stackrel} 		
\usepackage{tikz} 			
\usepackage{braket} 		
\usepackage{bm} 			
\usepackage{tensor} 		
\usepackage{slashed} 		
\usepackage{siunitx} 		
\usepackage{lastpage} 		
\usepackage{cite} 			
\usepackage[normalem]{ulem} 
\usepackage{fontawesome} 	
\usepackage{tocloft} 		
\usepackage{titlesec} 		
\usepackage{doi} 			
\usepackage{hyperref} 		
\usepackage[most]{tcolorbox} 					
\usepackage[nameinlink, capitalize]{cleveref} 	
\usepackage[nottoc, notlot, notlof]{tocbibind} 	
\usepackage[ruled, vlined]{algorithm2e} 		
\usepackage{makecell}
\usepackage[makeroom]{cancel}
\usepackage{feynmf}
\usepackage{aas_macros}          

\makeatletter
\def\BState{\State\hskip-\ALG@thistlm}
\makeatother

\makeatletter
\@ifundefined{pdfoutput}{}{\DeclareGraphicsRule{*}{mps}{*}{}}
\makeatother

\makeatletter
\DeclareRobustCommand*{\bfseries}{%
   \not@math@alphabet\bfseries\mathbf
   \fontseries\bfdefault\selectfont
   \boldmath
}
\makeatother

\hypersetup{
	pdftitle={},
	pdfauthor={Schosser et al.},
	colorlinks=true, 			
	linkcolor={red!50!black}, 	
	citecolor={blue!50!black}, 	
	urlcolor={blue!80!black} 	
} 

\DeclareSymbolFont{usualmathcal}{OMS}{cmsy}{m}{n}
\DeclareSymbolFontAlphabet{\mathcal}{usualmathcal}

\SetArgSty{textnormal}
\SetKwComment{Comment}{{\small\#}~}{}
\SetCommentSty{mycommfont}

\setitemize{itemsep=0pt, parsep=0pt} 				
\setenumerate{itemsep=0pt, parsep=0pt} 				
\setitemize{itemsep=2pt,topsep=2pt,parsep=0pt,partopsep=0pt,leftmargin=*}
\setenumerate{itemsep=0pt,topsep=2pt,parsep=0pt,partopsep=0pt,labelindent=3pt,leftmargin=*}
\usepackage{amsmath}
 
\usepackage{amsthm} 		
\theoremstyle{definition}

\definecolor{Rcolor}{HTML}{E99595}
\definecolor{Gcolor}{HTML}{C5E0B4}
\definecolor{Gcolor_light}{HTML}{F1F8ED}
\definecolor{Bcolor}{HTML}{9DC3E6}
\definecolor{Ycolor}{HTML}{FFE699}
\definecolor{Pcolor}{HTML}{C8B7E1}
\definecolor{Ycolor_light}{HTML}{FFF7DE}

\usetikzlibrary{arrows, arrows.meta, shapes, positioning}

\tikzstyle{arrow} = [thick,-{Latex[scale=1.0]}, line width=0.2mm, color=black]
\tikzstyle{loss} = [circle, thick, rounded corners=0.3ex, minimum width=1.5cm, minimum height=1cm, text centered, align=center, inner sep=0, fill=white, font=\large, draw]
\tikzstyle{net} = [double arrow, double arrow head extend=0cm, double arrow tip angle=130, shape border rotate=90, inner sep=0, align=center, minimum width=2.1cm, minimum height=2.3cm, fill=Bcolor, draw,font=\large]
\tikzstyle{net_black} = [net, minimum height=2.5cm, fill=black]
\tikzstyle{expr} = [rectangle, rounded corners=0.3ex, minimum width=2.6cm, minimum height=0.6cm, text centered, align=center, inner sep=0, fill=Ycolor, font=\large, draw]

\tikzstyle{small_cinn} = [double arrow, double arrow head extend=0cm, double arrow tip angle=130, inner sep=0, align=center, minimum width=1.1cm, minimum height=0.5cm, fill=Rcolor, draw]

\tikzstyle{transformer} = [rectangle, rounded corners, minimum width=6cm, minimum height=2.4cm, font=\large, fill=Gcolor_light, draw]

\tikzstyle{attention} = [rectangle, rounded corners=0.3ex, minimum width=5.5cm, minimum height=1.2cm, align=center, fill=Gcolor, draw, font=\large]

\tikzstyle{transformer_huge} = [rectangle, rounded corners, minimum width=8.5cm, minimum height=2.4cm, font=\large, fill=Gcolor_light, draw]

\tikzstyle{attention_huge} = [rectangle, rounded corners=0.3ex, minimum width=8cm, minimum height=1.2cm, align=center, fill=Gcolor, draw, font=\large]

\tikzstyle{txt_huge} = [align=center, font=\Huge, scale=2]
\tikzstyle{txt} = [align=center, font=\LARGE, minimum height=0.4cm]
 
\tcbset{
        enhanced,
        colback=gray!5!white,
        boxrule=0.1pt,
        colframe=gray!75!black,
        fonttitle=\bfseries,
        halign=flush left,
        width=0.90\linewidth,
       }

\DeclareSymbolFont{usualmathcal}{OMS}{cmsy}{m}{n}
\DeclareSymbolFontAlphabet{\mathcal}{usualmathcal}

\definecolor{red_cb}{HTML}{e41a1c}
\definecolor{blue_cb}{HTML}{377eb8}
\definecolor{green_cb}{HTML}{4daf4a}
\definecolor{purple_cb}{HTML}{984ea3}
\definecolor{orange_cb}{HTML}{ff7f00}

\definecolor{EmeraldGreen}{HTML}{1ea78d}
\definecolor{EnglishRed}{HTML}{b02427}
\hypersetup{colorlinks=true,urlcolor=EmeraldGreen,citecolor=EmeraldGreen,linkcolor=EnglishRed}

\newcommand{\eg}{\text{e.g.}\;}
\newcommand{\ie}{\text{i.e.}\;}

\newcommand{\rbb}{\mathbb{R}}

\newcommand\one{\leavevmode\hbox{\small1\normalsize\kern-.33em1}}

\newcommand{\softmax}{\operatorname{Softmax}}

\newcommand{\loss}{\mathcal{L}} 	
\newcommand{\normal}{\mathcal{N}} 	

\newcommand{\madgraph}{\textsc{MadGraph}\xspace}
\newcommand{\madagents}{\textsc{MadAgents}\xspace}

\newcommand{\arXiv}[2][]{%
	\ifthenelse{\equal{#1}{}}%
	{\href{http://arxiv.org/abs/#2}{arXiv:#2}}%
	{\href{http://arxiv.org/abs/#2}{arXiv:#2~[#1]}}}

\newcommand{\gev}{\text{GeV}}

\def\slashchar#1{\setbox0=\hbox{$#1$}           
   \dimen0=\wd0                                 
   \setbox1=\hbox{/} \dimen1=\wd1               
   \ifdim\dimen0>\dimen1                        
      \rlap{\hbox to \dimen0{\hfil/\hfil}}      
      #1                                        
   \else                                        
      \rlap{\hbox to \dimen1{\hfil$#1$\hfil}}   
      /                                         
   \fi}

\newcommand{\tikznode}[2]{%
\ifmmode%
\tikz[remember picture,baseline=(#1.base),inner sep=0pt] \node (#1) {$#2$};%
\else
\tikz[remember picture,baseline=(#1.base),inner sep=0pt] \node (#1) {#2};%
\fi}

\def\mathswitchr#1{\relax\ifmmode{\mathrm{#1}}\else$\mathrm{#1}$\xspace\fi}
\def\mathswitch#1{\relax\ifmmode#1\else$#1$\xspace\fi}

\graphicspath{{./figs/}}

\begin{document}


\vspace*{-2.5em}
\hfill{}
\vspace*{0.5em}

\begin{center}{\Large \textbf{
One Generator, Any Process: LLM-Conditioning for the LHC
}}\end{center}

\begin{center}
  Henning Bahl\textsuperscript{1},
  Tilman Plehn\textsuperscript{1,2},
  Daniel Schiller\textsuperscript{1}, and
  Thanush Sivagnanalingam\textsuperscript{1}
\end{center}

\begin{center}
{\bf 1} Institut f\"ur Theoretische Physik, Universit\"at Heidelberg, Germany\\
{\bf 2} Interdisciplinary Center for Scientific Computing (IWR), Universit\"at Heidelberg, Germany
\end{center}

\begin{center}
\today
\end{center}


\section*{Abstract}
{\bf Neural network training for LHC event generation should, ideally, benefit from common high-level patterns in different processes. We propose novel conditioning schemes for continuous parameters, process labels, and Feynman diagrams. We employ pre-trained LLMs as multi-modal foundation models to provide descriptive embeddings for an autoregressive transformer. With such high-level physics-inductive bias the generative networks converge faster, provide better result, and generalize to unseen processes.
}

\vspace{10pt}
\noindent\rule{\textwidth}{1pt}
\tableofcontents\thispagestyle{fancy}
\noindent\rule{\textwidth}{1pt}
\vspace{10pt}

\clearpage
\section{Introduction}

The volume of LHC data is growing rapidly, and modern machine learning (ML) is providing the methodology to target complex correlation using these large amounts of data~\cite{Butter:2022rso,Plehn:2022ftl}. Transformers with their exquisite expressivity provide the leading architectures~\cite{Mikuni:2021pou,Qu:2022mxj,DiBello:2022iwf,Finke:2023veq}, provided that we complement them with physics-specific inductive biases and data representations to optimize their performance for limited resources. Two complementary strategies in representation learning motivate   explicit~\cite{Brehmer:2024yqw,Favaro:2025pgz,Petitjean:2025zjf} or implicit~\cite{Breso-Pla:2026tlz} encoding of broken Lorentz symmetry in transformers. The latter is related to the idea of experimental foundation models trained on large collections of multi-modal detector or simulation data~\cite{Mikuni:2024qsr, Birk:2026udp, Mokhtar:2026gdb}. These foundation models aim to learn transferable representations that can be adapted to a wide range of downstream tasks. The extreme case of implicit representations is language models (LLMs), raising the question whether their pretrained embeddings can be exploited in physics~\cite{Heneka:2025fpe}.

In general, the choice of representation can determine whether a neural network generalizes beyond its training data. In LHC physics, we have converged on representations like 4-momenta, Lorentz or permutation invariants, and set-based jet or event encodings~\cite{Komiske:2018cqr,Qu:2019gqs,Bogatskiy:2020tje,Qu:2022mxj,Gong:2022lye,Bahl:2024gyt,Bahl:2025xvx}. Additional high-level information, for example, from theory predictions for scattering processes, has received comparatively little attention. 

An example for high-level physics information is processes that differ only by flavor labels, by an additional gluon~\cite{Butter:2024zbd}, by a single coupling exchange, or just share the same final state multiplicity. Transformers can exploit these high-level correlations through an appropriate embedding. Such an embedding has to cover information that cannot be accommodated as one-hot labels, like continuous masses and coupling, symbolic descriptions of initial and final states, or Feynman diagrams. Technically, such a representation should generalize to zero-shot predictions for processes that were not part of the training dataset.

Multi-modal transformers are the natural vehicle for general process representation. We benchmark pretrained LLM backbones~\cite{Heneka:2025fpe,2026arXiv260209670K} as the conditioning network of an autoregressive generative network for parton-level events. The LLM receives a description of the process as a one-hot label, text string of incoming and outgoing particles, list of Feynman-diagram edges, or rendered diagram images. We study the representation of continuous parameters such as a particle mass, as well as more general information encoded in Feynman diagrams. Conditional on previously generated outgoing 4-momenta a conditional flow matching (CFM) head generates the next-particle momentum distribution for our proof-of-concept implementation.

First, we consider Drell-Yan production with a varying $Z$ mass and compare embeddings of the mass value as a one-hot bin, via a small mass-embedding network, or as text. We then evaluate the network for mass values never seen during training. Second, we start with a catalog of leading-order $2\to 2$ and $2\to 3$ Standard Model processes and generate events for processes not seen before. For both settings, descriptive embeddings provide a physics inductive bias that one-hot labels lack. They converge faster, generalize to unseen processes, generate substantially better event samples, and are easier to fine-tune even for processes with four or more final-state particles. 

In Sec.~\ref{sec:setup} we describe the architecture, embedding strategies, training datasets, and training and evaluation protocols. In Sec.~\ref{sec:continous_conditioning} we present a study targeting the Drell-Yan process comparing different conditioning schemes for the continuous $s$-channel $Z$ mass. In Sec.~\ref{sec:process_conditioning} we target process conditioning and provide an in-depth discussion of text-based, image-based, and graph-based embedding. Technical details and additional information are collected in the Appendices.

\section{Conditioned event generation}
\label{sec:setup}

As an application of multi-modal process conditioning, we use the generation of the outgoing momenta for parton-level LHC processes at tree-level~\cite{Butter:2019cae,Butter:2021csz}. For each process, we order the final-state particles and represent each event by the 4-momentum sequence $(k_1, \dots, k_n)$. We factorize the joint distribution in an auto-regressive manner, 
\begin{align} 
  p(k_1, \dots, k_n | \text{process}) = \prod_{i=1}^n p_\theta(k_i | k_{[1 : i-1]}, \; \text{process})
  \qquad \text{with} \qquad 
  k_{[1 : i-1]} \equiv k_1, \dots, k_{i-1} \; ,
    \label{eq:autoregressive_gen}
\end{align}
and encode the conditional probabilities with a transformer.

\subsection*{Network architecture}

Our autoregressive transformer uses a causal attention mask coupled to a conditional flow matching (CFM) head.  The transformer provides the conditioning of the previous momenta together with the process information, and the CFM head encodes the conditional probability of the next momentum. Its learned velocity field transports random numbers to the target distribution,
\begin{align} \label{eq:cfm_ode}
    &\frac{d \tilde{k}_i (\tau)}{d \tau} = v_\theta(\tilde{k}_i, \tau, k_{[1 : i-1]}, \; \text{process}) \notag \\
    & \text{with} \quad \tilde{k}_i \sim \begin{cases}
        \text{noise} & \tau=0 \\
        p(k_i | k_{[1 : i-1]}, \; \text{process}) & \tau=1 \; .
    \end{cases}
\end{align}
Our CFM for LHC events assumes straight trajectories from the noise to the target distribution and is trained on the corresponding continuity equation~\cite{Plehn:2022ftl,Butter:2023fov},
\begin{align}
    &\loss = \left< \left(v(k_i, \epsilon) - v_\theta(\tilde{k}_i, \tau, k_{[1 : i-1]}, \; \text{process}) \right)^2 \right>_{p(k_i | k_{[1 : i-1]}, \; \text{process}), \text{noise}(\epsilon), U_{[0, 1]}(\tau)} \notag\\
    & \text{where} \quad v \equiv k_i - \epsilon \quad \text{and} \quad \tilde{k}_i \equiv \tau k_i + (1 - \tau) \epsilon  \; .
\end{align}
The CFM head is described in more detail in App.~\ref{app:CFM}. Our architecture can be straightforwardly adapted to regression or classification heads.

We parametrize the conditional and predicted 4-momenta as
\begin{align}
    k_i = \left( \log (m^2 + \epsilon), \log (p_T + \epsilon), \eta, \phi \right) \, ,
\end{align}
with a small regulator $\epsilon$. For on-shell final-state particles, we omit $\log (m^2 + \epsilon)$ from the generation target and fix the external masses instead. The CFM samples standardized $\log p_T$ and $\eta$ from unit Gaussians and $\phi$ from a uniform distribution. To simplify the periodic boundaries, we represent $\phi$ as $\left(\cos\phi, \sin\phi\right)$ and augment the data by randomly shifting $\phi$ in the same way for all 4-momenta of each event.

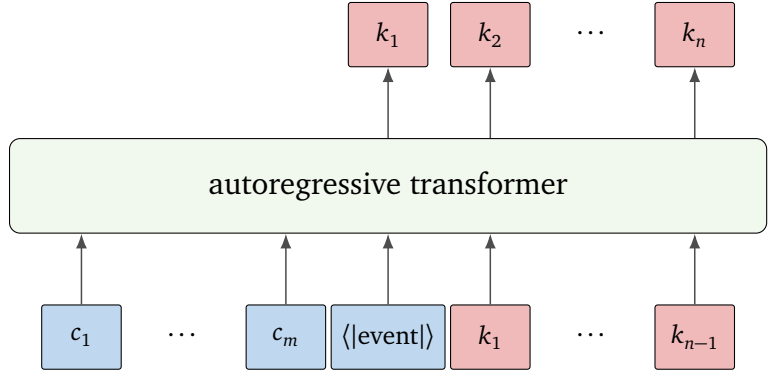
\begin{figure}[t]
    \centering
    \begin{tikzpicture}[x=1.35cm, y=1.0cm]
\tikzset{
  cell/.style={
    draw, rounded corners=0.2ex,
    minimum width=1.05cm,
    minimum height=0.85cm,
    font=\small, align=center
  },
  dotcell/.style={
    minimum width=1.05cm,
    minimum height=0.85cm,
    font=\small, align=center
  },
  inarrow/.style={arrow, color=black!70},
  outarrow/.style={arrow, color=black!70}
}

\def\yin{0.0}
\def\ytr{2.0}
\def\yout{4.0}

\node (c1)   [cell, fill=Bcolor!65] at (0,\yin) {$c_1$};
\node (cd)   [dotcell]              at (1,\yin) {$\cdots$};
\node (cm)   [cell, fill=Bcolor!65] at (2,\yin) {$c_m$};

\node (ev)   [cell, fill=Bcolor!65] at (3,\yin) {$\langle|\mathrm{event}|\rangle$};

\node (p1i)  [cell, fill=Rcolor!65] at (4,\yin) {$k_1$};
\node (pdi)  [dotcell]              at (5,\yin) {$\cdots$};
\node (pnmi) [cell, fill=Rcolor!65] at (6,\yin) {$k_{n-1}$};

\node (tr) [transformer, minimum width=10.0cm, minimum height=1.25cm]
  at (3,\ytr) {autoregressive transformer};

\node (p1o) [cell, fill=Rcolor!65] at (3,\yout) {$k_1$};
\node (p2o) [cell, fill=Rcolor!65] at (4,\yout) {$k_2$};
\node (pdo) [dotcell]              at (5,\yout) {$\cdots$};
\node (pno) [cell, fill=Rcolor!65] at (6,\yout) {$k_n$};

\foreach \name in {c1,cm,ev,p1i,pnmi}{
  \draw[inarrow] (\name.north) -- (\name.north |- tr.south);
}

\foreach \name in {p1o,p2o,pno}{
  \draw[outarrow] (\name.south |- tr.north) -- (\name.south);
}

\end{tikzpicture}
    \caption{Autoregressive generative architecture. The prefix tokens $c_1, \dots, c_m, \langle|\text{event}|\rangle$ condition the generation of the outgoing momenta $k_1, \dots, k_n$.}
    \label{fig:arch_overview}
\end{figure}

We sketch our network architecture in Fig.~\ref{fig:arch_overview}. We decompose the input into the prefix tokens, $c_1, \dots, c_m, \langle|\text{event}|\rangle$, and momenta $k_1, \dots, k_{n-1}$.  We distinguish the event token from the other prefix tokens because the first momentum $k_1$ is generated over this token. 

\subsection*{Process embedding}

The prefix tokens encode the process information using multiple modalities. Different aspects may be useful for learning efficient representations, from continuous mass or coupling values to discrete particle IDs. Our autoregressive transformers map a sequence of token vectors, $(x_1, x_2, \dots, x_n) \in \rbb^{d \times n}$ to an output sequence,
\begin{align}
\begin{pmatrix} x_1 \\ x_2 \\ \vdots \\ x_n \end{pmatrix} \in \rbb^{n \times d}
\; \mapsto \; 
\begin{pmatrix} f_\theta(x_1) \\ f_\theta(x_1, x_2) \\ \vdots \\ f_\theta(x_1, x_2, \dots, x_n) \end{pmatrix} \in \rbb^{n \times d} \; .
\end{align}
$\rbb^{d}$ denotes the latent space of the transformer. Each modality requires a dedicated network to embed the data in the latent space, to then be used in the input sequence. To simplify the notation, we write the embedding networks of all modalities as $E_\theta$, implicitly assuming that they differ from modality to modality.

For process information, the embedding vectors are used as prefix tokens. There would also be the option to aggregate multiple vectors into one, for instance via addition. We describe five generic modalities and their embeddings:
\begin{enumerate}
    \item \textbf{categorical labels}, up to the limiting case where each process defines its own category. For each label $l \in \{ l_1, \dots, l_n\}$, a latent embedding vector is learned,
\begin{align}
    l \mapsto E_\theta(l) \in \rbb^d.
\end{align}

\item \textbf{numerical values}, such as a continuous mass or coupling. The raw value is passed through a small connector network that maps it directly into the latent space,
\begin{align}
    m \mapsto E_\theta(m) \in \rbb^d \; .
\end{align}

\item \textbf{text}, for which we use pretrained LLMs as the autoregressive transformer. The tokenizer splits the text into a sequence of discrete tokens to be treated as categorical labels. The pretrained LLM provides embedding vectors for those tokens,
\begin{align}
    \text{text} \mapsto (t_1, \dots, t_n) \;, \quad t_i \mapsto E_\theta(t_i) \in \rbb^d \; .
\end{align}

\item \textbf{images}, for which we use pretrained multi-modal LLMs as the autoregressive transformer. Each image is embedded into the latent space, concretely a sequence of visual tokens,
\begin{align}
    \text{image} \in \rbb^{\text{channels} \times \text{width} \times \text{height}} \mapsto E_\theta(\text{image}) = (v_1, \dots, v_n) \in \rbb^{d \times n} \; .
\end{align}
The number of visual tokens depends on the resolution and the embedding network typically consists of a vision transformer and a convolutional network. Unlike text embedding, the sequence of visual tokens is jointly embedded. Text and visual tokens are aligned with the latent space of the multi-modal LLM.

\item \textbf{graphs}, for which we use a dedicated graph network to obtain an embedding which is then passed to the backbone transformer. Instead of parsing a serialized text or image description with a pretrained LLM, we directly embed Feynman diagrams~\cite{Mitchell:2022yxp} and replace the pretrained LLM by a small transformer trained from scratch. The graph nodes are the external particles and the vertices, the edges are the external legs and internal propagators, carrying a particle identity. A small graph transformer with multi-head self-attention over the nodes, into which the edge particle identities enter through learned terms~\cite{shaw2018selfattention}, encodes one diagram into a single latent vector,
\begin{align}
    \text{graph} \mapsto E_\theta(\text{graph}) \in \rbb^d \; .
\end{align}
The graph attention is described in detail in App.~\ref{app:Feynman_diagram_input}.
\end{enumerate}

\noindent
Due to the large scale of the pretrained Qwen2.5-0.5B~\cite{yang2024qwen2} and Qwen3-VL-2B~\cite{qwen3vl2025} LLMs, we partially fine-tune both. First, we duplicate the backbone weights, freeze one copy and train the other, following the same general idea as Ref.~\cite{dong2024internlm}. We use the frozen pre-aligned weights for the textual and visual tokens and the trainable weights for the remaining modalities, such as the particle momenta and the event token. The causal attention mask allows us to cache the key-value activations of the frozen modalities, as long as no trainable modality precedes them in the input sequence. We use a hydragen-style~\cite{juravsky2024hydragen} attention decomposition. Our text and image input sequences are small enough for the attention computation to be at most comparable to the feed-forward evaluation, rendering this conditional almost free of compute cost. Further details can be found in App.~\ref{app:technical_details}.

Following Ref.~\cite{Heneka:2025fpe}, we wrap the input sequence with the chat template:
\begin{tcolorbox}[colback=gray!5!white,colframe=gray!75!black]
    \texttt{<|im\_start|>} system\\
    \texttt{<|im\_end|>}\\
    \texttt{<|im\_start|>} user\\
    \{process conditioning\} \texttt{<|im\_end|>}\\
    \texttt{<|im\_start|>} assistant\\
    \texttt{<|event|>} $k_1$ $\dots$ $k_{n-1}$ \texttt{<|im\_end|>}
\end{tcolorbox} 
\noindent
The tokens \texttt{<|$\cdot$|>} are special LLM tokens indicating system, user, and assistant messages. We only add the token \texttt{<|event|>}, with an embedding learned during training. Further details about the LLMs and their training can be found in the App.~\ref{app:technical_details}.

\subsection*{Momentum embedding}

We map each 4-momentum $k_i$ into the transformer latent space, now with a simple linear layer
\begin{align}
    k_i  \mapsto E_\theta(k_i) \in \rbb^d \; .
\end{align}
In addition, we can encode the particle ID as a categorical label and combine its  embedding vector with the 4-momentum as
\begin{align}
    (k_i, \text{pID}) \mapsto E_\theta(k_i) + E_\theta'(\text{pID}) \in \rbb^d  \; ,
\end{align}
where $E_\theta(k_i)$ and $E_\theta'(\text{pID})$ are independent from each other. Since we fix the order of outgoing particle IDs, this embedding does not add new information, but it may help to learn an efficient process representation.

\clearpage
\section{Parameter conditioning}
\label{sec:continous_conditioning}

\begin{figure}[b!]
    \includegraphics[width=0.495\linewidth, page=1]{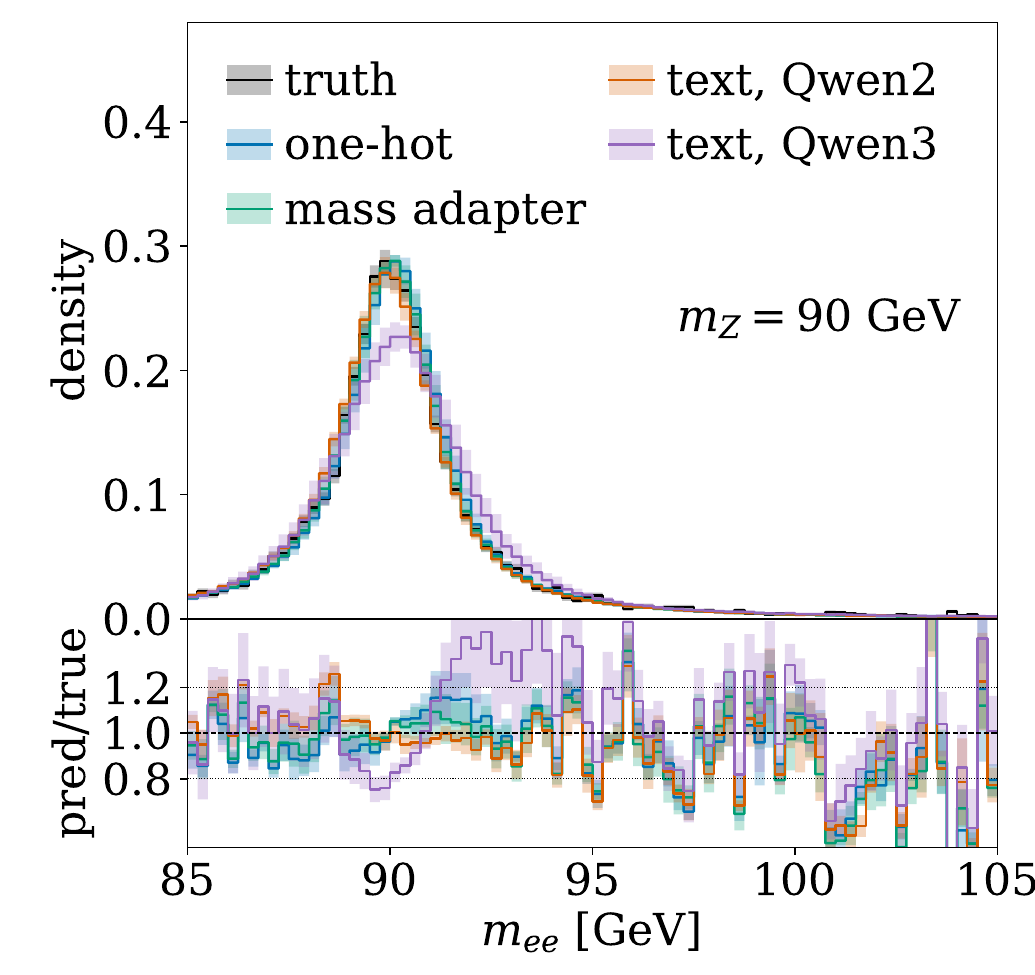}
    \hfill
    \includegraphics[width=0.495\linewidth, page=1]{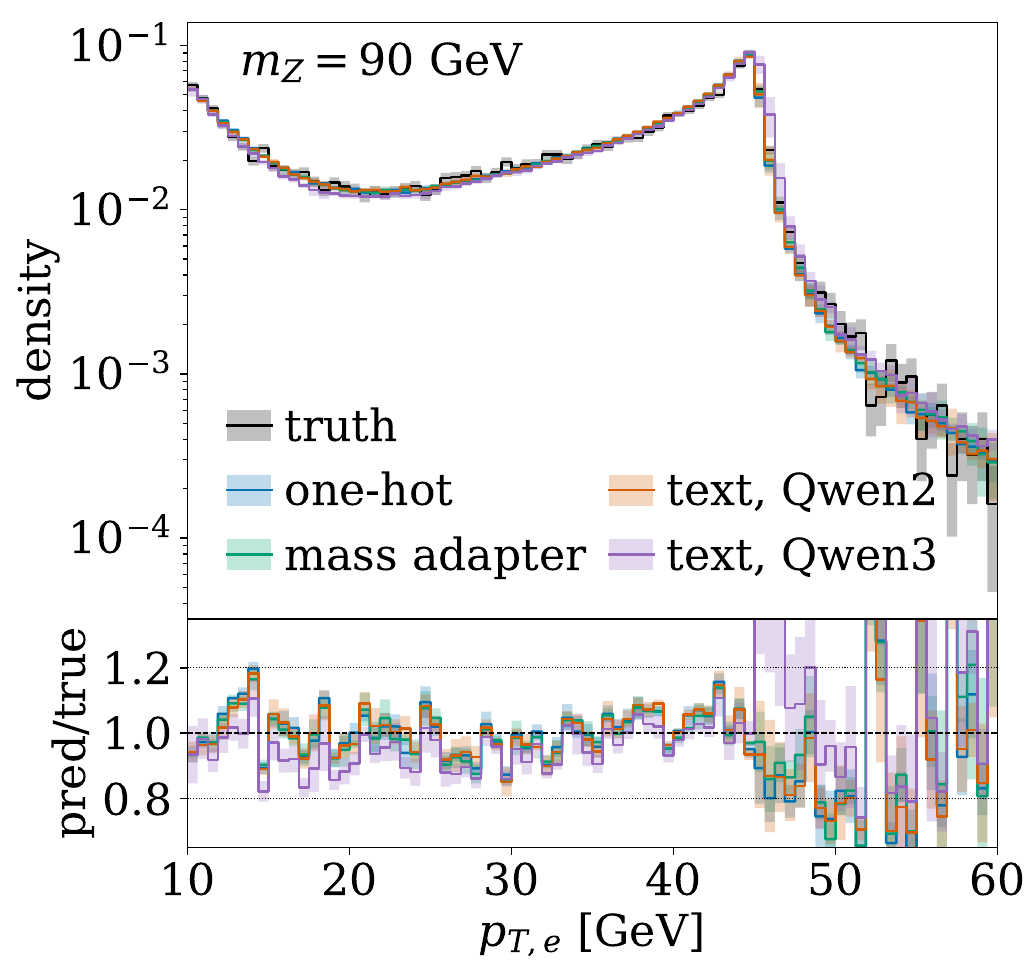}\\
    \includegraphics[width=0.495\linewidth, page=1]{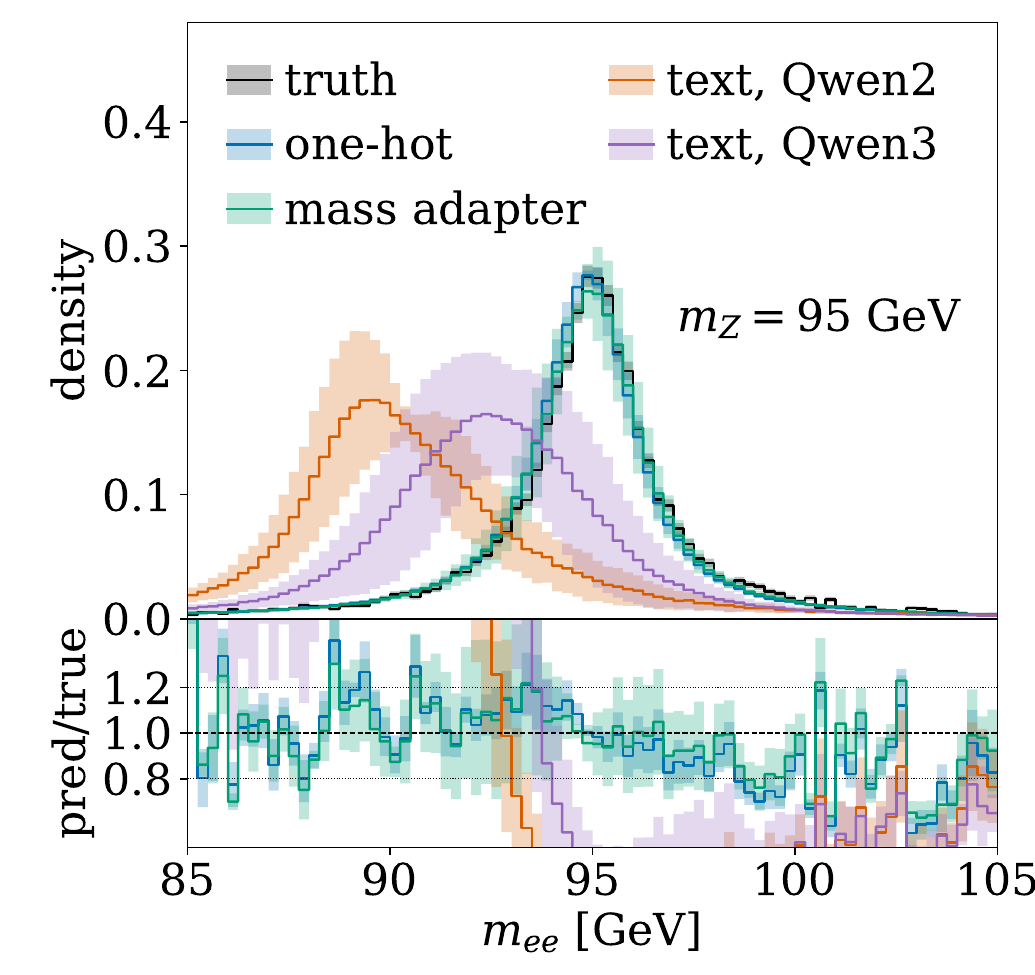}
    \hfill
    \includegraphics[width=0.495\linewidth, page=1]{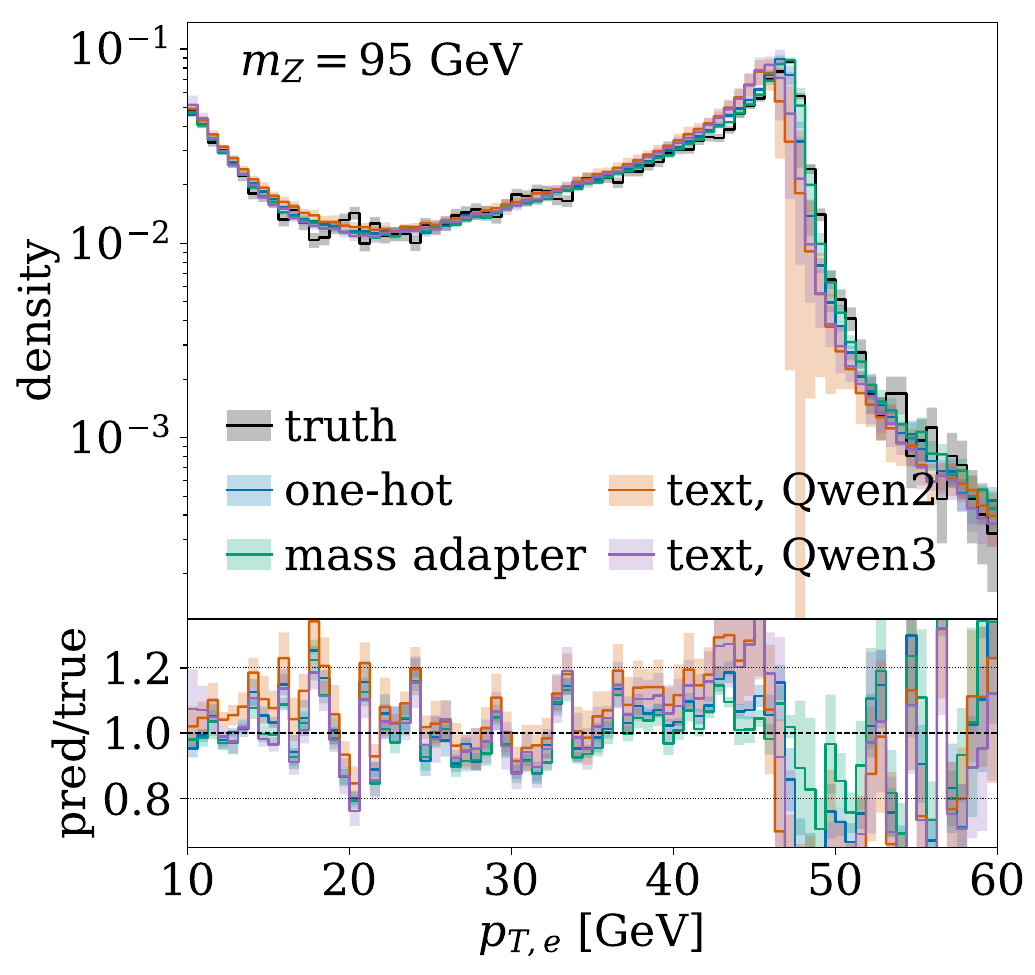}
    \caption{Left: Drell-Yan invariant mass distribution for $m_Z = 90\,\gev$ (upper) and $m_Z = 95\,\gev$ (lower). The colored bands indicate the standard deviation over five independent runs. Right: same for the positron transverse momentum.}
    \label{fig:DY}
\end{figure}

First, we investigate conditioning on a continuous physics parameter for Drell-Yan production 
\begin{align}
  q\bar q\to e^+ e^-
  \qquad \text{conditional on} \quad
  m_Z \; .
\end{align}
We generate all events at leading-order (LO) parton level with \madgraph~\cite{Alwall:2014hca} and basic generation cuts only. In the training dataset, we condition the $Z$ mass~\cite{Baldi:2016fzo} in the two windows $[80,90]\,\gev$ and $[100,110]\,\gev$ in 1~GeV steps. For each of the 22 values, we use 70k events for training. We then generate events for $m_Z = 95\,\gev$ to test the process representation and its generalization. We compare three mass conditioning schemes:
\begin{enumerate}
    \item \textbf{one-hot:} unique ID for each $m_Z$ value as a discrete mass label. 
    \item \textbf{text:} $m_Z$ value provided as text input.
    \item \textbf{mass adapter:} small MLP taking $m_Z$ as input and providing an  embedding.
\end{enumerate}
The embedding vectors of the one-hot encoding and the mass adapter are added to the event token. As the transformer backbone, we use the Qwen2 model with 0.5B parameters. For the case of text input, we test the Qwen3 model with 2B parameters in addition. Each conditioning scheme is trained independently five times. Both LLMs are partially fine-tuned, with the details and hyperparameter listed in App.~\ref{app:technical_details}. Without this dedicated embedding, conditional LHC generators have been developed for event unfolding~\cite{Bellagente:2019uyp,Bellagente:2020piv}. Other established conditionings include theory nuisance parameters~\cite{Butter:2021csz}, phase space mappings~\cite{Heimel:2022wyj,Heimel:2023ngj} helicities~\cite{Bothmann:2025lwg,Bothmann:2026dar}, FKS sectors~\cite{DeCrescenzo:2026tsp}, or a combination of both~\cite{Janssen:2025zke}.

In the upper panels of Fig.~\ref{fig:DY}, we show the di-lepton invariant mass and  positon transverse momentum distributions for $m_Z = 90\,\gev$. The distributions from all encodings are similar and close to the truth, as $m_Z = 90\,\gev$ is covered by the training data. The only noticeable deviation appears for Qwen3, with a reduced density around the $Z$ peak. This likely reflects the larger number of trainable parameters in the Qwen3 backbone, which may require stronger regularization or longer training to match the performance of Qwen2.

In the lower panels of Fig.~\ref{fig:DY}, we interpolate to $m_Z = 95\,\gev$. The one-hot encoding does not have an embedding vector for this $Z$ mass. In Fig.~\ref{fig:param_random_embd} of App.~\ref{app:supplementary_results}, we display the corresponding distributions with random embedding vectors. To investigate whether the one-hot process representation covers the hold-out processes, we finetune the embedding vector of the one-hot run. Now, the one-hot embedding and the mass adapter recover the mass peak, whereas the text input does not, with the larger Qwen3 performing better than the smaller Qwen2. Similarly, for the positron transverse momentum the finetuned one-hot encoding and the mass adapter match the true distribution, while text conditioning deviates near the $Z$ peak.

\begin{figure}[t]
    \centering
    \includegraphics[width=0.495\linewidth, page=1]{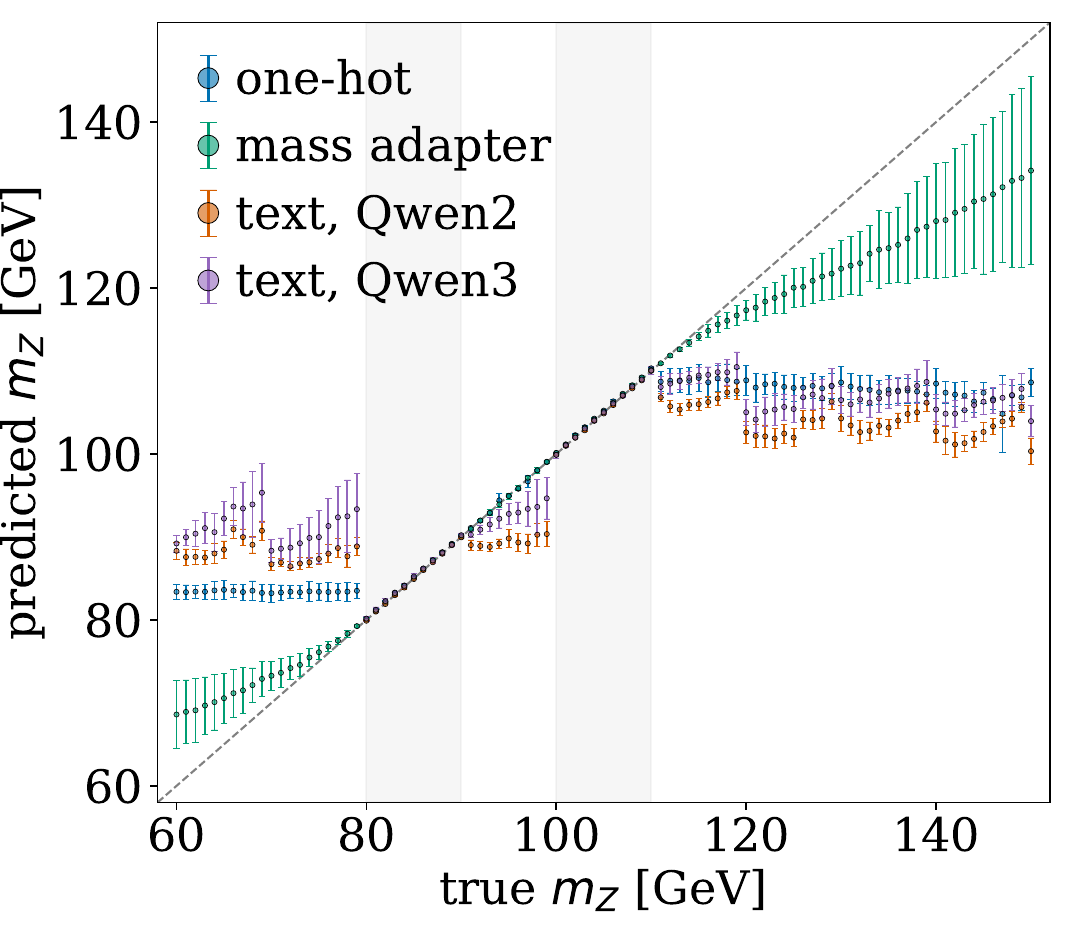}
    \caption{Predicted $Z$ mass, extracted from the generated invariant mass peak, as a function of the true $Z$ mass. The shaded regions indicate the in-training masses. The uncertainty bars indicate the standard deviation over five independent runs.}
    \label{fig:DY_pred_vs_true}
\end{figure}

For Fig.~\ref{fig:DY_pred_vs_true}, we scan a wider range of $Z$ masses and identify the predicted mass as the peak position in the invariant mass. We show the mean and standard deviation  for different true $Z$ masses. All embeddings perform well on the training masses, but differ significantly on the unseen $Z$ masses. The one-hot representation only covers the interpolation region. The mass-adapter outperforms the finetuned one-hot encoding, interpolates well, but gradually degrades when extrapolating to smaller and larger masses. The text embedding fails to generalize even into the interpolation region. The predicted masses exhibit a zigzag shape which we suspect to originate from the network mainly attending to the last digit, since each digit is tokenized into a single token. Qwen3 outperforms the smaller Qwen2 backbone in the interpolation region, which we suspect comes from a better latent representation of digits and numbers, consistent with simple arithmetic emerging only at sufficiently large model scale~\cite{wei2022emergent}.

\section{Process conditioning}
\label{sec:process_conditioning}

Next, we condition on the process type, where the process label is a discrete but highly structured object. Our process catalog contains over a thousand distinct processes, tied to one another by flavor symmetries, shared interaction vertices, and common diagram topologies. A good conditioning scheme should make this structure accessible to the generative network, to allow it to generate processes not seen during training. We compare process embeddings of increasing descriptiveness:
\begin{enumerate}
    \item \textbf{one-hot:} unique process id, for which a dedicated embedding vector is trained. The momenta come with an additional particle type embedding.
    \item \textbf{in-out:} \madgraph-inspired string for in- and outgoing particles, \eg, `g g > t t\scalebox{0.6}[1]{\textasciitilde} h'.
    \item \textbf{in-interm-out:} an extended string with intermediate particles, \eg, `g g > [g, t] > t t\scalebox{0.6}[1]{\textasciitilde} h'.
    \item \textbf{edge-list:} in addition to the in-interm-out description, we provide it a text representation of the Feynman diagrams. This edge list encodes which vertices are connected by which propagator and provides a lossless, one-to-one encoding of the Feynman diagram. Details can be found in App.~\ref{app:Feynman_diagram_input}.
    \item \textbf{image:} supplementing in-interm-out with a graphical representation of each Feynman diagram, which we take from \madgraph.
\end{enumerate}

\begin{table}[b!]
    \centering
    \begin{small}
    \begin{tabular}{l l}
        \toprule
        Hold-out category & probed symmetry/coupling \\
        \midrule
        $s\bar s$-initiated processes                & degeneracy with $u\bar u$, $d\bar d$ \\
        muonic final states from $b\bar b$           & $e/\mu$ universality \\
        $g+s$ and $b+s$ scattering                   & $d\leftrightarrow s$ flavour generalization \\
        $c+\bar u$ scattering                        & isospin generalization \\
        same-sign light quarks (e.g.\ $ds\to ds$)    & light-flavour combinatorics \\
        multi-Higgs ($HH$, $HHH$, $ZHH$, $WHH$)      & Higgs self-coupling \\
        triple gauge-boson production                & quartic/neutral-boson topologies \\
        $H\tau\tau$-vertex processes                 & $\tau$ Yukawa coupling \\
        $gg\to t\bar t + X$                          & top Yukawa coupling \\
        $gg\to W qq'$                                & charged current from pure-QCD initial state \\
        same-sign $cb$/$\bar c\bar b$ scattering     & heavy-flavour combinatorics \\
        \bottomrule
    \end{tabular}
    \end{small}
    \caption{Overview of the $139$ hold-out processes used to probe generalization.}
    \label{tab:holdout}
\end{table}

\subsection*{Networks and training}

With the help of \madagents~\cite{Plehn:2026gxv}, we generate events for the complete catalog of $2\to 2$ and $2\to 3$ tree-level Standard Model processes at the LHC, $254$ processes for $2\to 2$ and $1237$ for $2\to 3$ scattering, $1491$ processes in total. All events are generated in the Standard Model with a diagonal CKM matrix, massless light fermions, and generation cuts $p_T > 10$ GeV and $|\eta| < 5$. We train on $1352$ processes with $10^5$ events per process, and hold out the remaining $139$ to test generalization. The hold-out processes contain processes which are similar to processes in the training data, probing flavor and lepton universality present in the training distribution, and processes probing interaction vertices and topologies absent from training.  In Tab.~\ref{tab:holdout}, we summarize the categories. The full list can be found in App.~\ref{app:supplementary_results}.

We generally use Qwen2 to reduce the backbone dependence of the runs, except for the image embedding which requires the larger Qwen3 model. For each process embedding scheme, we train five independent runs. Analogously to the previous section, the LLMs are partially finetuned, where the text and image tokens use the pretrained LLM weights, while we train another set of weights for the new modalities. Details and hyperparameter are shown in App.~\ref{app:technical_details}. 

\begin{figure}[t]
    \centering
    \includegraphics[width=0.7\linewidth]{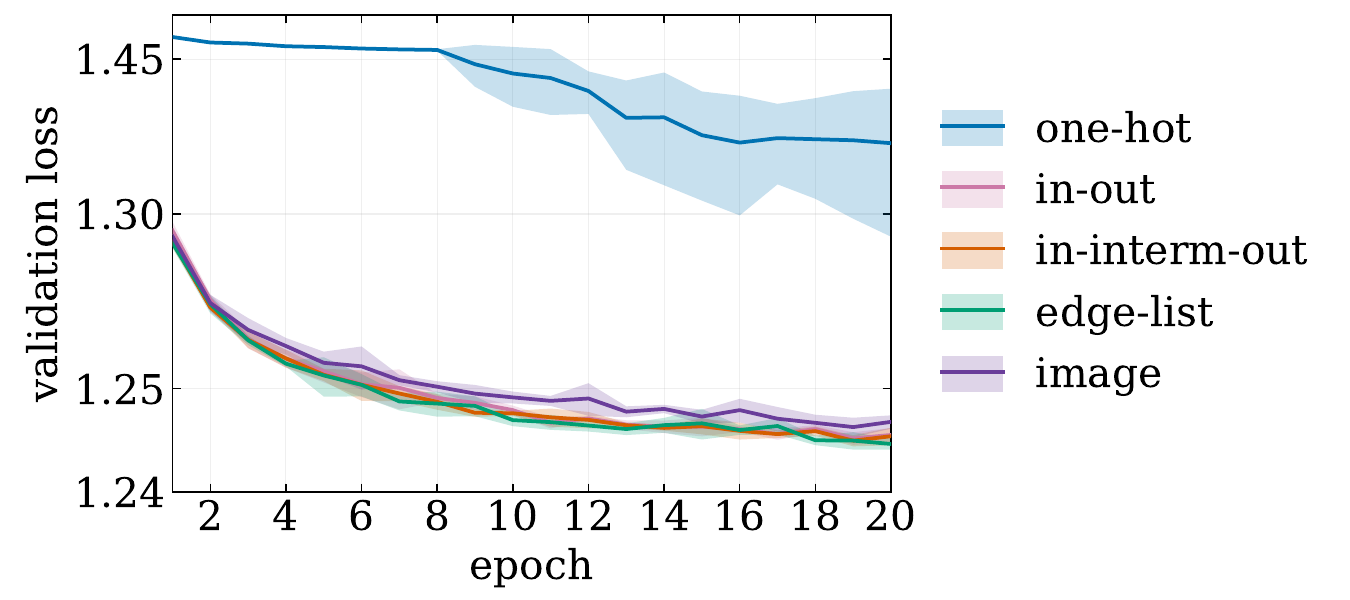}
    \caption{Loss comparison over $20$ epochs for different conditioning mechanisms. The colored bands indicate the standard deviation over five independent runs.}
    \label{fig:loss_curves}
\end{figure}

We show the loss curves for the various configurations in Fig.~\ref{fig:loss_curves}. Providing discriminative information between the different processes into the context yields an inductive bias, which leads to much faster convergence. Within the training time of 20 epochs, the one-hot encoding does not reach the performance of the other conditioning schemes.

\subsection*{Process extrapolation}

\begin{figure}[t]
    \centering
    \includegraphics[width=0.99\linewidth]{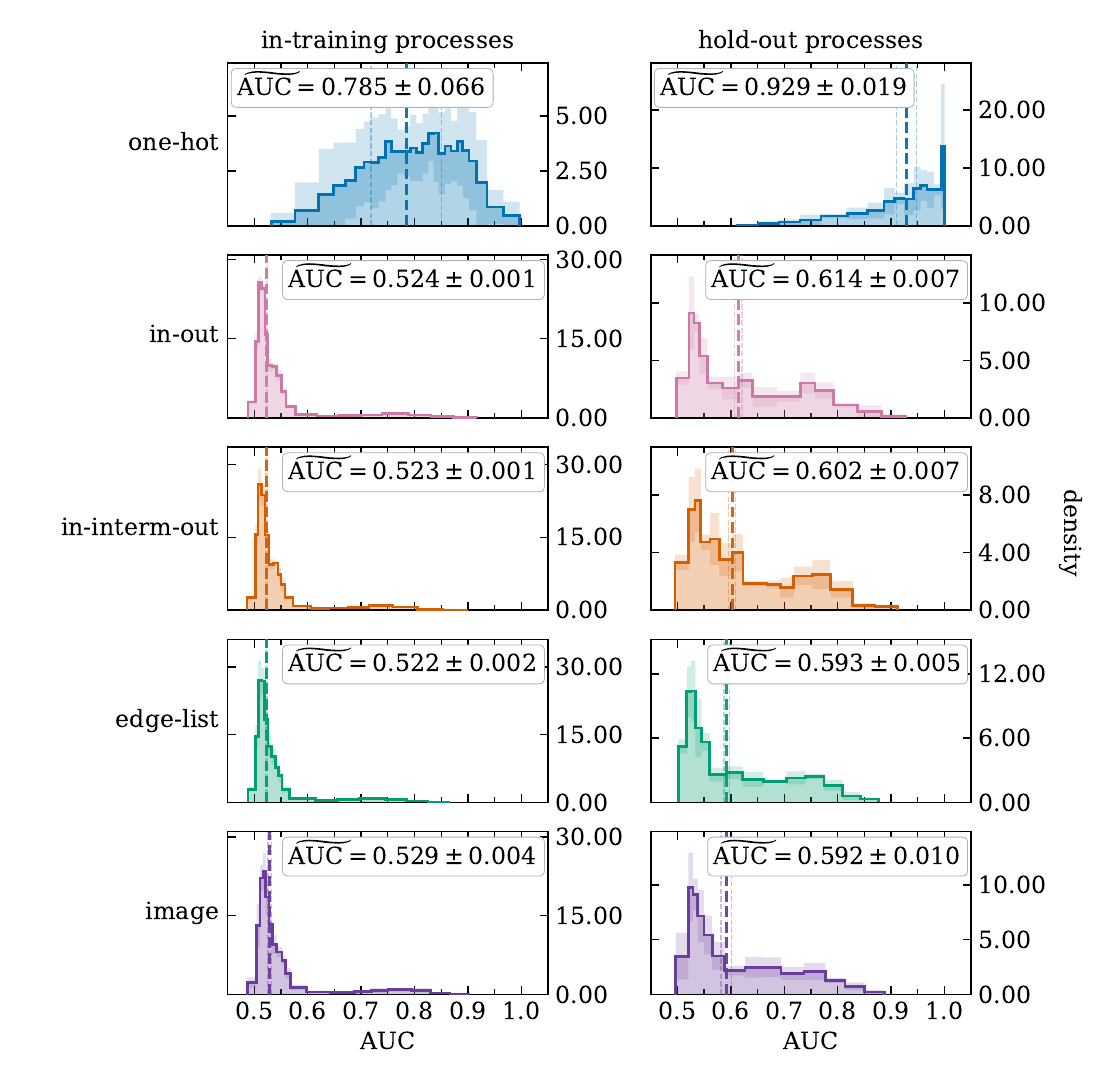}
    \caption{AUCs for different conditioning schemes. The dashed line indicates the median AUC, with uncertainties from averaging over five independent runs.}
    \label{fig:proc_encoding_comparison}
\end{figure}

In Fig.~\ref{fig:proc_encoding_comparison}, we compare the results of different process embeddings. For each embedding and process, we train a classifier to distinguish the generated from the true events. The classifier is described in App.~\ref{app:technical_details}. We histogram the AUC values for the 1352 training in the left column and 139 hold-out processes in the right column. For each embedding, we indicate the median $\widetilde{\text{AUC}}$.

For the training processes the one-hot encoding fails with $\widetilde{\text{AUC}} \gtrsim 0.8$, whereas the text and image process encodings perform well, $\widetilde{\text{AUC}} \sim 0.5~...~0.65$.  Different backbones do not have a significant impact. Providing information about the Feynman diagrams does not lead to a significant improvement for the training processes.

When extrapolating to the hold-out processes, the one-hot encoding fails again. For the text and image embeddings the classifier scores $\widetilde{\text{AUC}}  \sim 0.6$ are only marginally worse than the values from the training processes. Starting from only the incoming and outgoing particles, information about the intermediate particles leads to an $\widetilde{\text{AUC}}$ improvement by $\sim 0.01$, explicit information about the Feynman diagrams of another improvement by $\sim 0.01$.

\begin{table}[b!]
    \centering
    \begin{small} \begin{tabular}{l|cc|cc}
        \toprule
        & \multicolumn{2}{c|}{training processes}
        & \multicolumn{2}{c}{hold-out processes} \\
        & median AUC & mean AUC
        & median AUC & mean AUC \\
        \midrule
        in-out        & $0.524(1)$ & $0.557(2)$ & $0.614(7)$  & $0.637(4)$ \\
        in-interm-out & $0.523(1)$ & $0.552(5)$ & $0.602(7)$  & $0.630(8)$ \\
        edge-list     & $0.522(2)$ & $0.548(4)$ & $0.593(5)$  & $0.621(5)$ \\
        image         & $0.529(4)$ & $0.563(6)$ & $0.592(10)$ & $0.622(6)$ \\
        \bottomrule
    \end{tabular} \end{small}
    \caption{Performance of different configurations. The quoted uncertainties represent the standard deviation averaging over five runs.}
    \label{tab:overview}
\end{table}

We present a summary of the performance measures in Tab.~\ref{tab:overview}. The median and mean AUC values are successively improved when we provide additional information. One-hot encoding is insufficient, and information about the incoming and outgoing particles is needed. Additional information about the intermediate particles improves the performance slightly for the training processes and substantially for the hold-out processes. Using edge lists or images representing the Feynman diagrams further improves performance, in particular for the hold-out processes. Further AUC overviews --- restricting the comparison to the $2\to3$ processes, repeating it for the Qwen3 backbone, and showing the AUC of every individual process --- are collected in App.~\ref{app:supplementary_results}.

\subsection*{Process kinematics}

\begin{figure}[t]
    \includegraphics[width=0.495\linewidth, page=2]{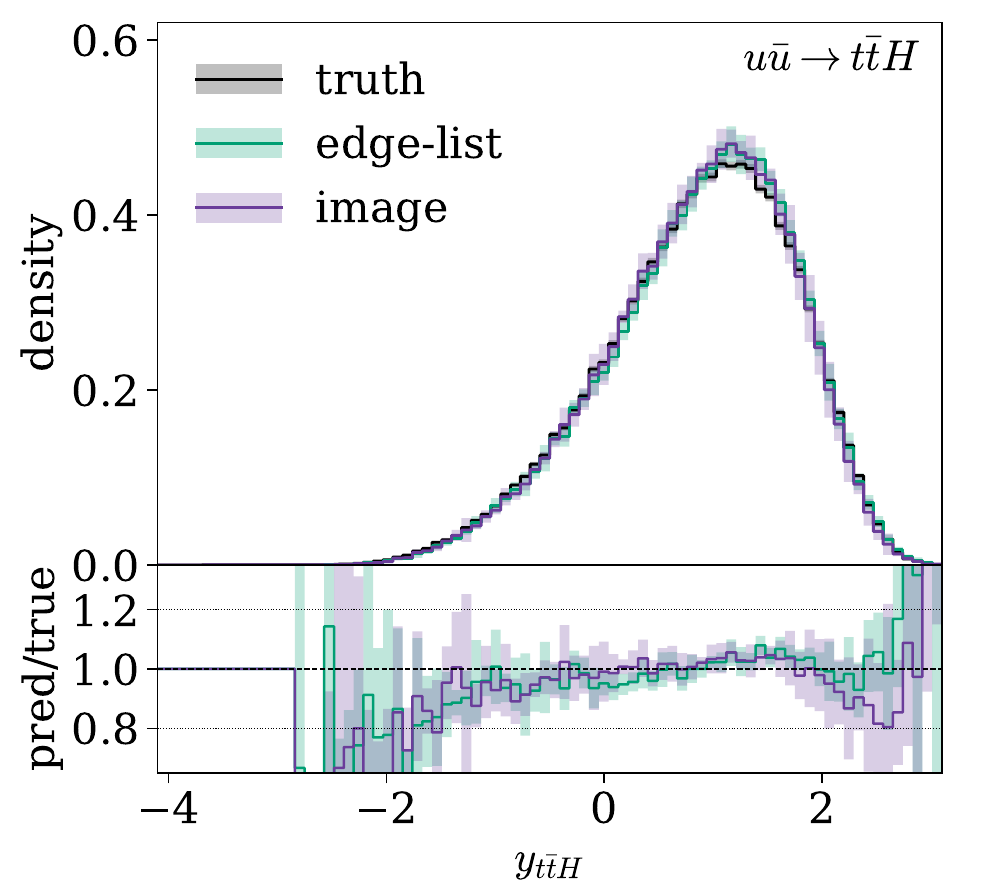}
    \includegraphics[width=0.495\linewidth, page=2]{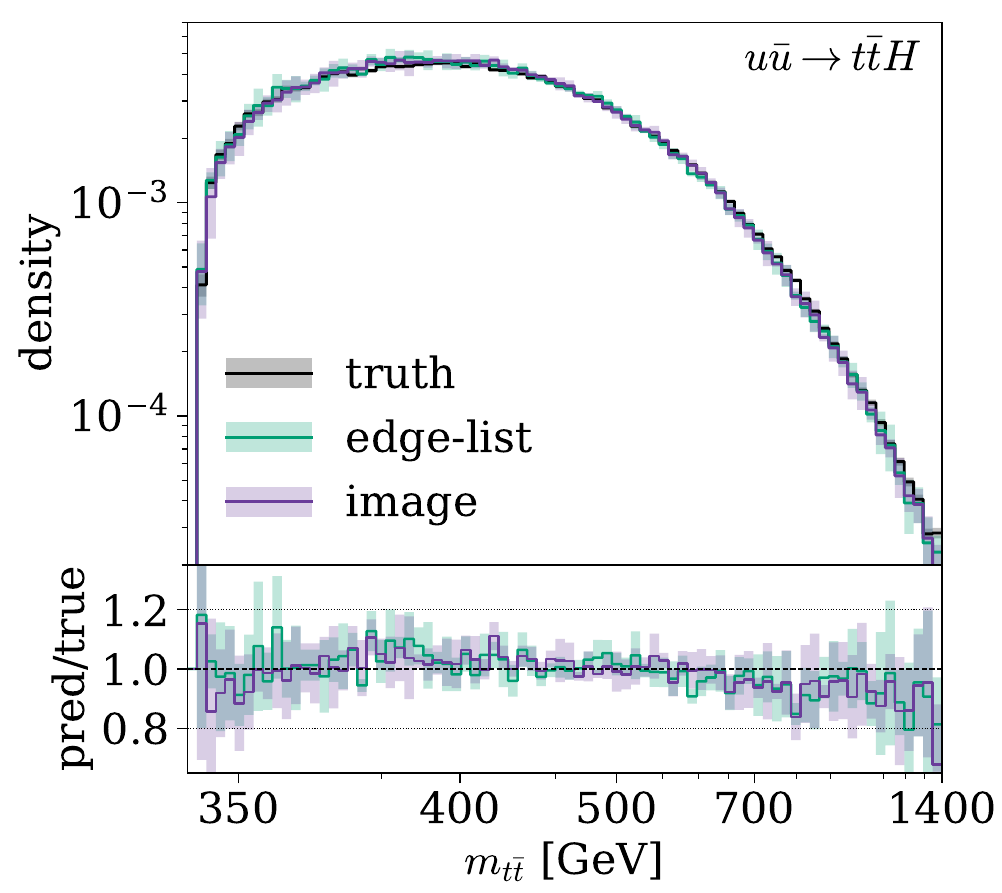}\\
    \includegraphics[width=0.495\linewidth, page=2]{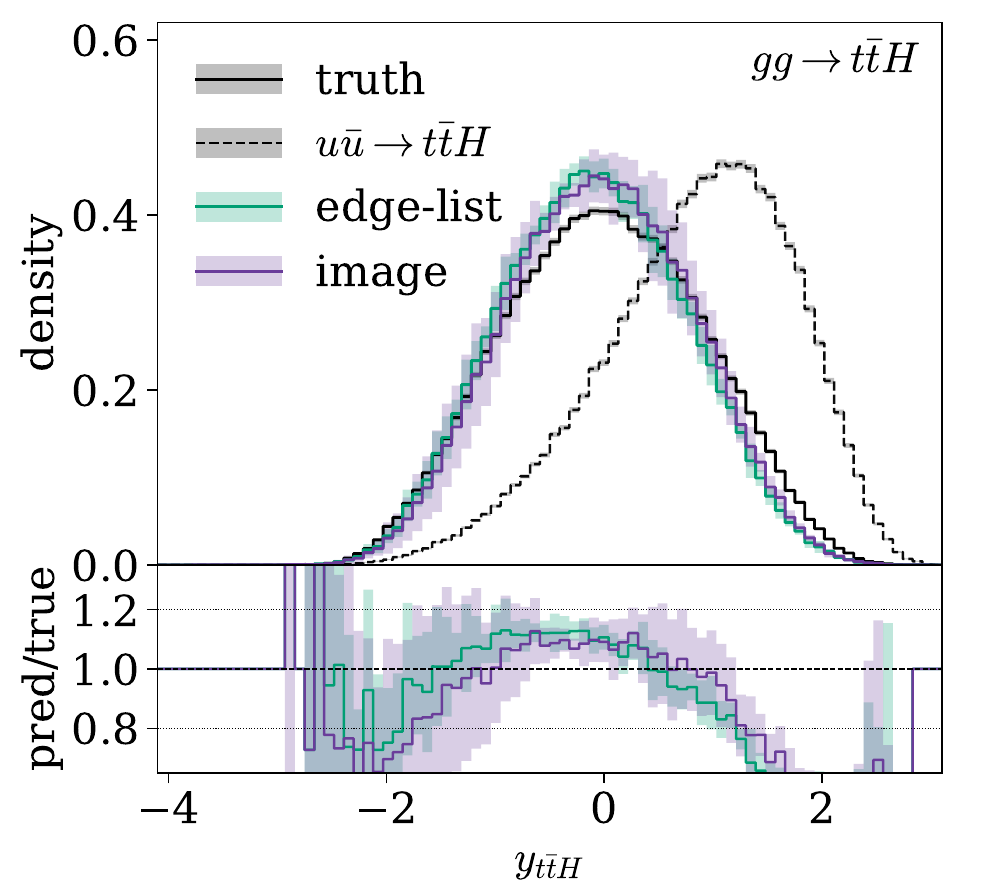}
    \includegraphics[width=0.495\linewidth, page=2]{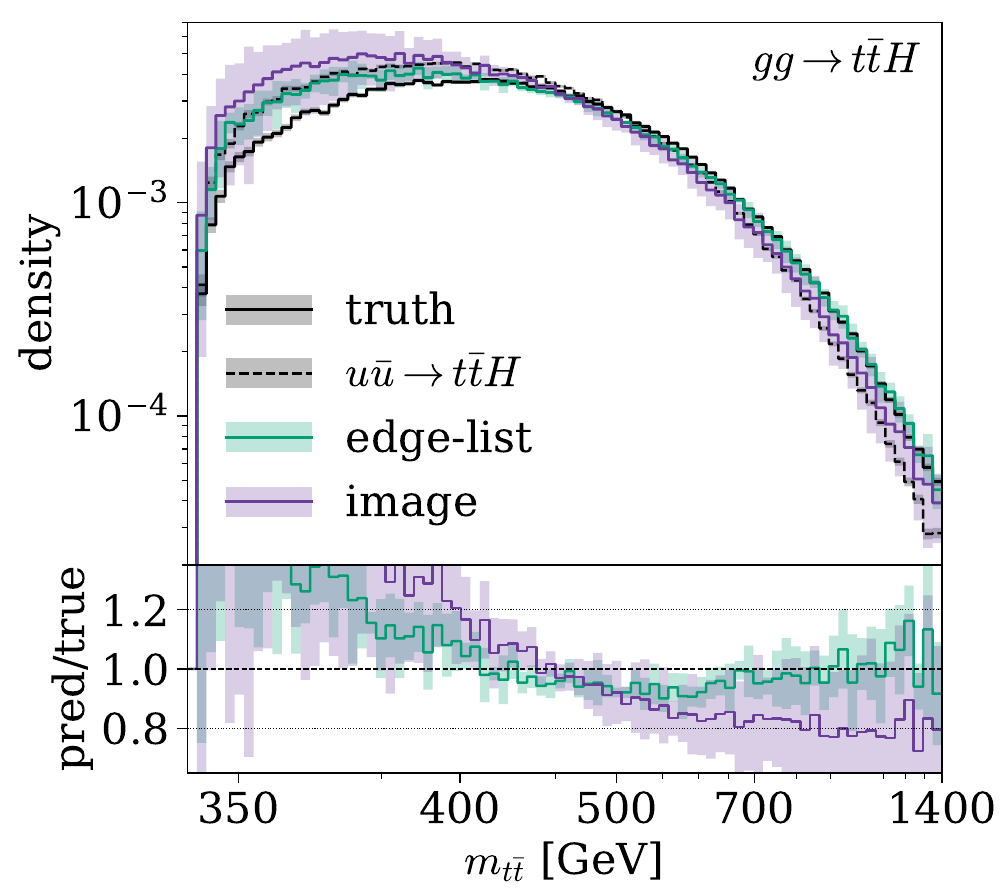}

    \caption{Kinematics for $u\bar{u}\to t\bar{t}H$ (in-training) and $gg\to t\bar{t}H$ (hold-out) production. We show the rapidity of the $t\bar{t}H$ system (left) and $m_{t\bar{t}}$ (right). The histograms depict the mean and standard deviation of 5 independent runs. In the lower panels we also show the $u\bar{u}\to t\bar{t}$ distributions for comparison.}
    \label{fig:uuvsggcomparison}
\end{figure}

For the example processes
\begin{align}
u\bar{u} &\to t \bar{t} H &\quad &\text{(in-training)} \notag \\
gg &\to t \bar{t} H       &\quad &\text{(hold-out)}
\end{align}
we investigate the event generation performance for one-hot, edge-list, and image runs in detail. Starting with the $u\bar{u}$ channel, part of the training dataset, we show in the upper panels of Fig.~\ref{fig:uuvsggcomparison} the rapidity of the $t\bar t H$ system on and $m_{t\bar t}$. All embeddings, with the exception of the generally insufficient one-hot encoding match the truth distributions.

The gluon-induced channel, shown in the lower panels of Fig.~\ref{fig:uuvsggcomparison}, is not part of the training dataset. The generative network has to deduce the kinematic distributions from the process embedding, combined with the training process $u\bar{u} \to t\bar t H$. However, the gluon parton density behaves differently than the quark-antiquark parton densities, and the $gg \to t\bar t H$ matrix element has a different color and spin structure.

In contrast to the $u\bar{u}$ training process, the rapidity distribution for the $gg$ channel has to be symmetric around zero. All embeddings produce the symmetric distributions and do not reproduce the behavior of the $u\bar u\to t\bar tH$ training channel.

The invariant mass of the top pair is sensitive to the matrix element itself. The edge-list embedding performs best in the high-mass tail, where the difference to the $u\bar u \to t\bar t H$ process is mainly due to parton densities. In the threshold region, where the changes from the matrix element are most visible, the image and text embedding of the Feynman diagrams both struggle to learn the shift from the $u\bar u$ channel and provide a distribution similar to the $u\bar u$ channel in the training data. This indicates that the transformer did not account for these subtleties, even though it could in principle learn the relevant coupling structure from in-training processes like $gg\to t\bar t$, $g b\to tWH$, and $u d\to t b H$. However, this is a much harder task.

\subsection*{Long-training limit}

\begin{figure}[t]
    \centering
    \includegraphics[width=0.99\linewidth]{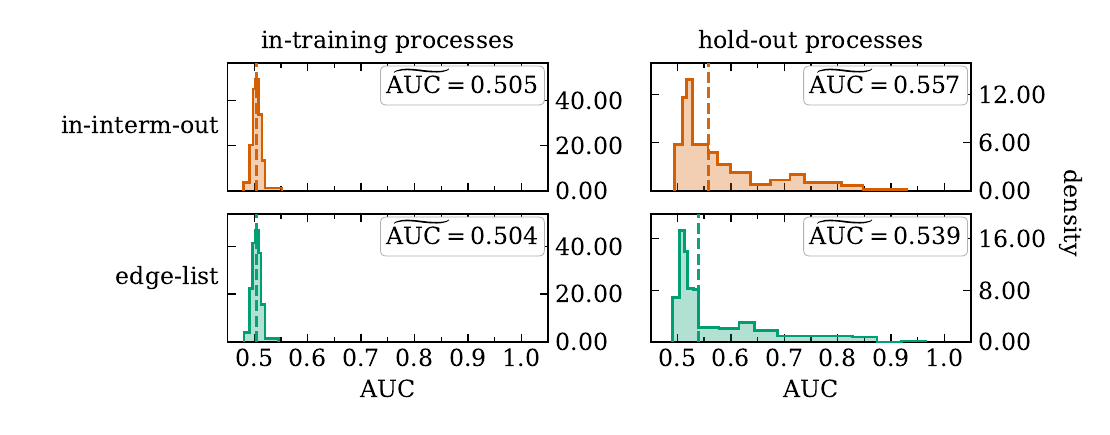}
    \caption{AUCs for different conditioning schemes trained for 120 epochs.}
    \label{fig:auc_long}
\end{figure}

To test the performance limit of our new setup, we continue training the two best in-interm-out and edge-list networks for 100 more epochs, \ie 120 epochs in total. The detailed training setup is explained in App.~\ref{app:technical_details} and the corresponding validation losses over the full 120 epochs are shown in App.~\ref{app:supplementary_results}.

We see in Fig.~\ref{fig:auc_long} that the AUC values for the training and hold-out processes decrease to the point that the classifiers can hardly distinguish true and generated events. However, for the hold-out processes, there remains a long tail towards higher AUC values. For our $m(t\bar t)$ distribution from the $gg\to t\bar tH$ hold-out process, we do not observe any improvement over the results shown above. This suggests that more training information is needed, for example, a set of $2\to 4$ processes.

\subsection*{Non-LLM graph embedding}

So far, our conditioning relies on a pretrained LLM to parse a serialized text or image description of a graph-like object. The graph embedding from Sec.~\ref{sec:setup} encodes each Feynman diagram with a dedicated graph network and lets us replace the pretrained LLM by a small transformer trained from scratch. A process is described by one or several diagrams, and we turn their per-diagram readouts into prefix tokens in two ways:
\begin{enumerate}
\item in the pooled variant, the diagrams are aggregated by attention into a single prefix token per process;
\item in the separate variant, each diagram contributes its own prefix token, and the aggregation is left to the self-attention of the backbone. 
\end{enumerate}
We also consider a permutation-invariant set encoder over the external particles, each embedded through its particle type and its incoming/outgoing direction, and attention-pooled into one prefix token. This in-out encoder carries the same information as the in-out text string of Sec.~\ref{sec:process_conditioning} and isolates how much of the performance is due to the graph structure rather than the external legs alone.

\begin{figure}[t]
    \includegraphics[width=0.99\linewidth]{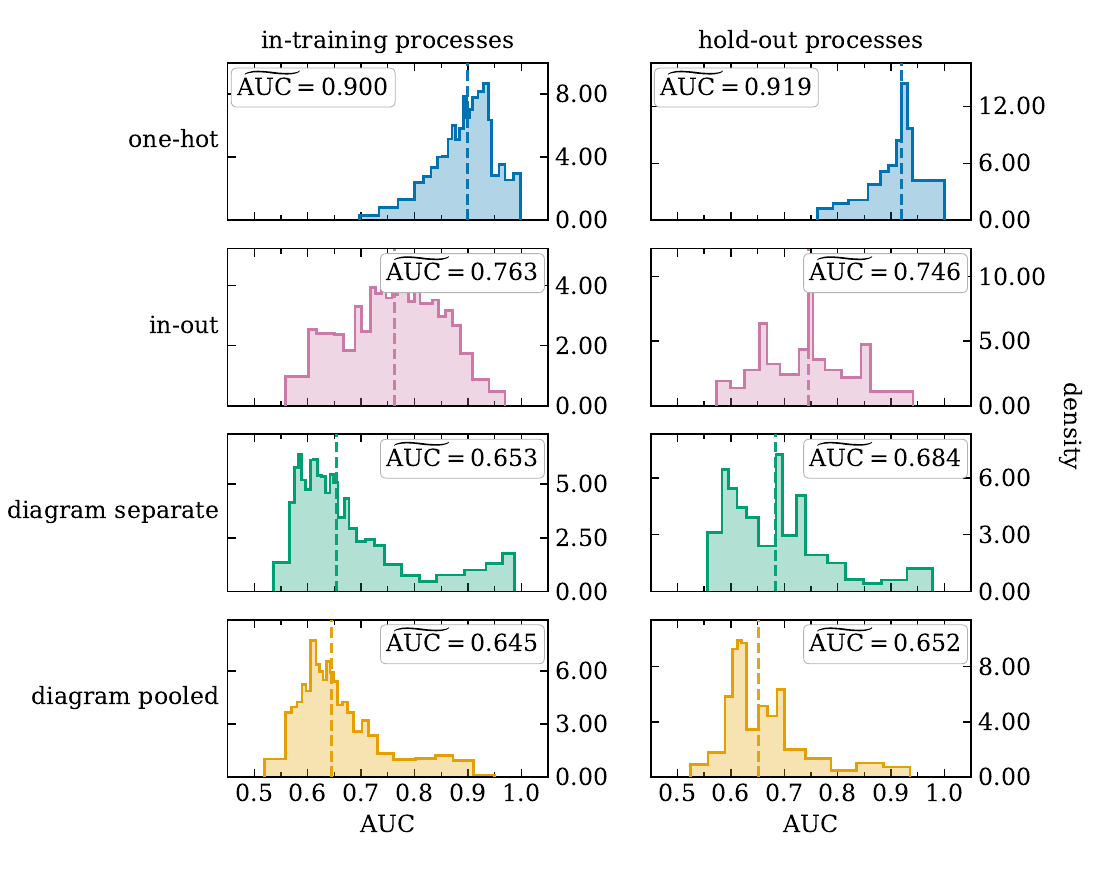}
    \caption{AUCs for different conditioning schemes for a non-LLM transformer backbone. The dashed line indicates the median AUC, with uncertainties from averaging over five independent runs.}
    \label{fig:graph_embedding_comparison}
\end{figure}

In Fig.~\ref{fig:graph_embedding_comparison}, we compare these graph embeddings  with the same small from-scratch transformer backbone against the one-hot encoding baseline. As before, we histogram the per-process classifier AUCs for the 1352 training processes (left) and the 139 hold-out processes (right). All runs are trained for 80 epochs. The one-hot encoding results are again poor with a median $\widetilde{\text{AUC}} \sim 0.90$ on training and hold-out processes. The in-out encoder substantially improves this to $\widetilde{\text{AUC}} \sim 0.75$, and adding the full diagram structure gives $\widetilde{\text{AUC}} \sim 0.65$. Pooling the diagrams into a single token performs marginally better than keeping them separate. 

Crucially, the graph embeddings generalize to the hold-out processes with hardly any performance loss. The hold-out medians track the training medians closely, indicating that the graph encoder captures process structure that transfers to unseen processes. Although the from-scratch transformer backbone size is much smaller than our LLM backbones, implying faster training and inference time, the generative performance is clearly worse.

\subsection*{Finetuning}

\begin{figure}[t]
    \centering
    \includegraphics[width=0.5\linewidth]{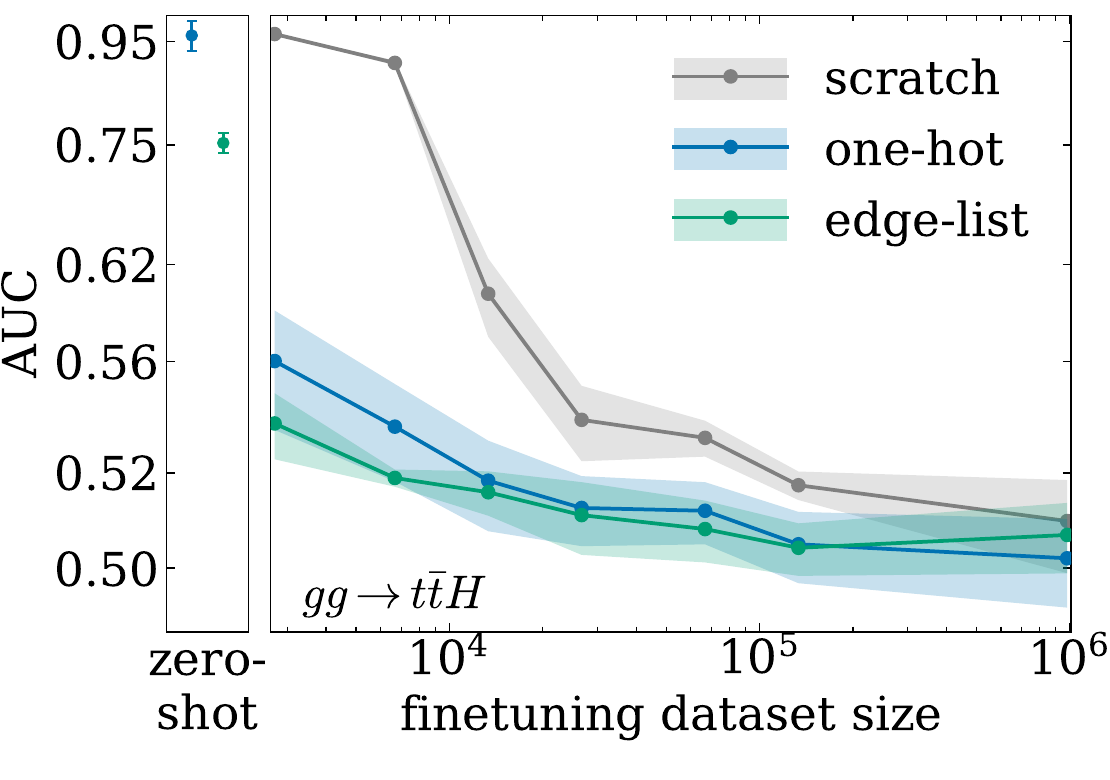}
    \caption{Classifier AUC for $gg \to t\bar t H$ as a function of the number of finetuning events, for the one-hot and edge-list conditioning compared to training from scratch. The left panel shows the zero-shot AUC of the pretrained networks. Lines and bands are the mean and standard deviation over five independent runs.}
    \label{fig:finetune_ggttH_auc}
\end{figure}

While generating events in a zero-shot fashion is an amusing challenge, the more relevant question for the LHC is how efficiently a pretrained network can be finetuned to a single hold-out process with a small number of events. We probe this finetuning for $(i)$ data efficiency; $(ii)$ extrapolation to more final-state particles; and $(iii)$ longer pretraining.

We pretrain on the full training dataset including $u \bar{u} \to t\bar t H$ events, and then fine-tune on gluon-induced events, and compare against the same Qwen2 backbone with a one-hot label, trained from scratch solely on the $gg\to t\bar t H$ process. Across all dataset sizes, we keep the batch size and the number of epochs fixed and use a cosine-annealing learning-rate schedule, such that a larger finetuning dataset translates into more optimizer steps per epoch. Additional training details can be found in App.~\ref{app:technical_details}.

\begin{figure}[b!]
    \includegraphics[width=0.495\linewidth, page=2]{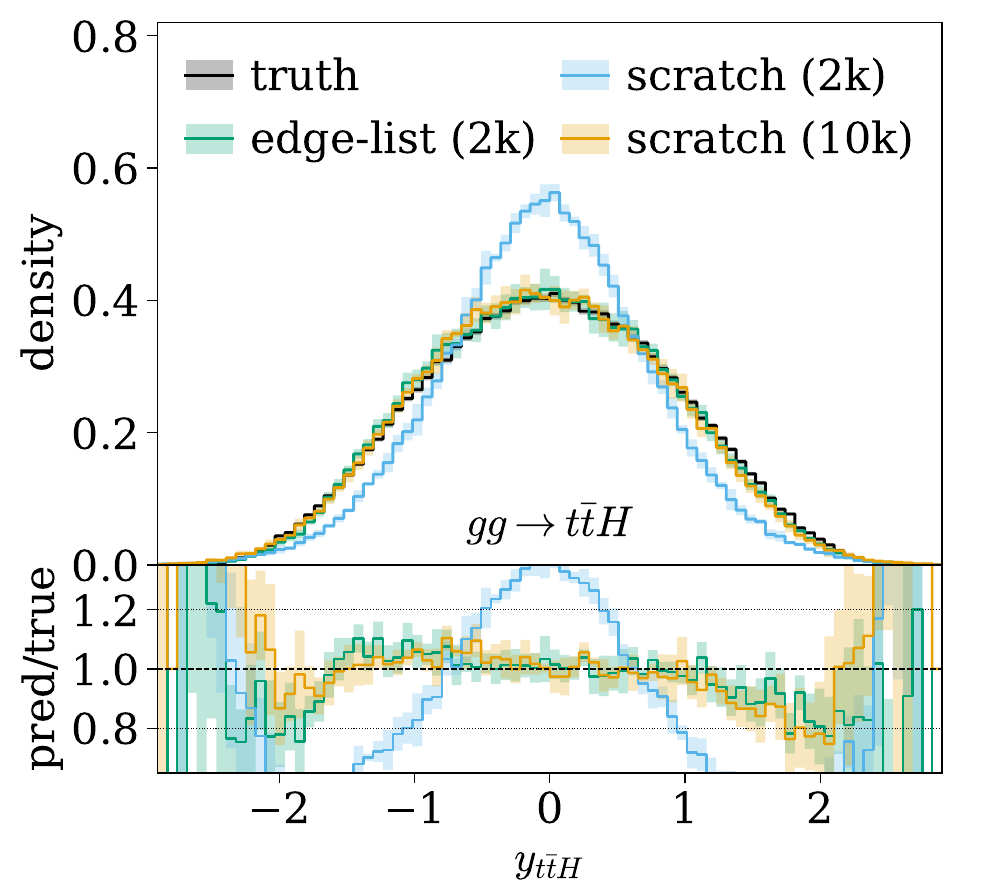}
    \includegraphics[width=0.495\linewidth, page=2]{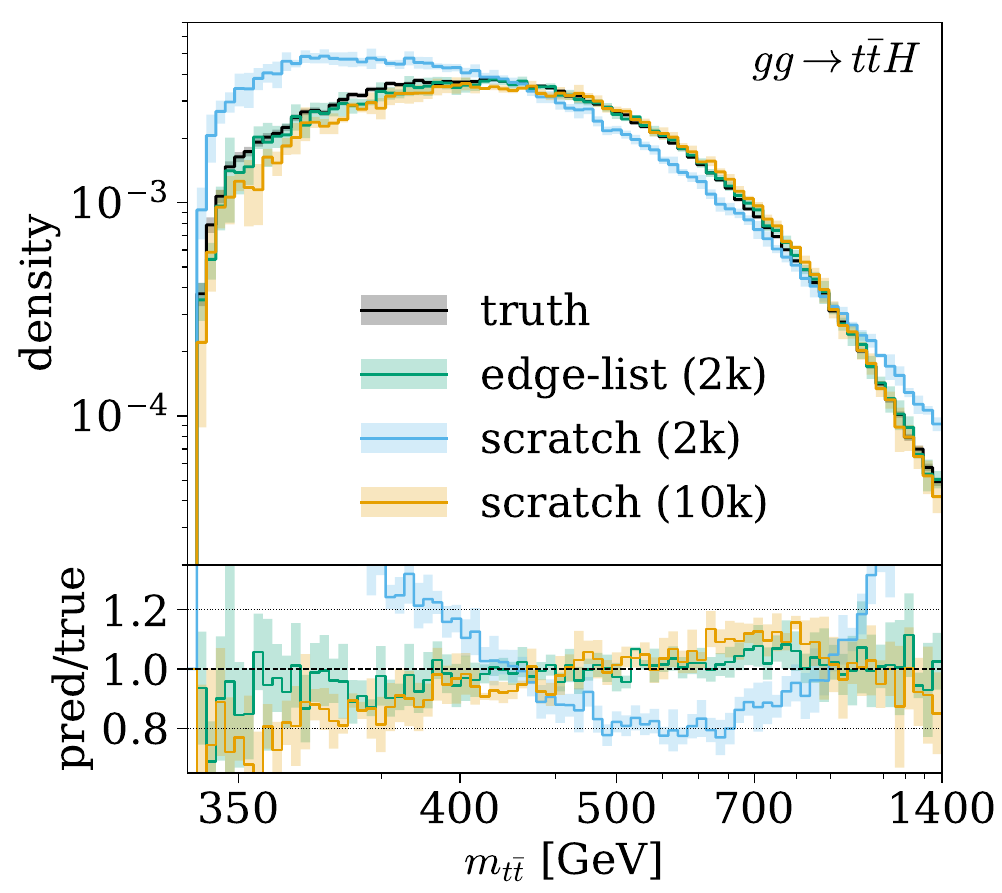}
    \caption{Rapidity of the $t\bar t H$ system (left) and $m_{t\bar t}$ (right) for $gg \to t\bar t H$, comparing the edge-list network finetuned on $2048$ events to the network trained from scratch on $2048$ and $10000$ events. Histograms show the mean and standard deviation over five independent runs, with the ratio to the truth in the lower panels.}
    \label{fig:finetune_ggttH_obs}
\end{figure}

In Fig.~\ref{fig:finetune_ggttH_auc}, we show the classifier AUC as a function of the number of finetuning events. Trained from scratch, the network only approaches the optimal $0.5$ for $10^6$ events. Finetuning the pretrained network reaches a comparable AUC with $2048$ events for the one-hot and for the edge-list conditioning. The performance gap persists for larger training samples. As discussed above, zero-shooting using the edge-list embedding is already much closer to the truth than the one-hot label. After finetuning, both converge to comparable performance, with the edge-list embedding retaining a small advantage in the low-statistics regime.

In Fig.~\ref{fig:finetune_ggttH_obs}, we compare the ($t\bar t H$) rapidity and $m_{t\bar t}$ distributions from the edge-list network finetuned on $2048$ events to  the from-scratch baseline at $2048$ and $10000$ events. The finetuned network reproduces the correct distributions within uncertainties. For 2k fine-tuning events, the from-scratch network instead produces a too narrow, over-peaked rapidity distribution and distorts $m_{t\bar t}$, overshooting the threshold region and undershooting the high-mass tail. For $10^5$ fine-tuning events, the distributions are significantly closer to the truth distribution but deviations are still slightly larger than for the pretrained model.

\subsection*{Final state multiplicity}

\begin{figure}[t]
    \includegraphics[width=0.495\linewidth]{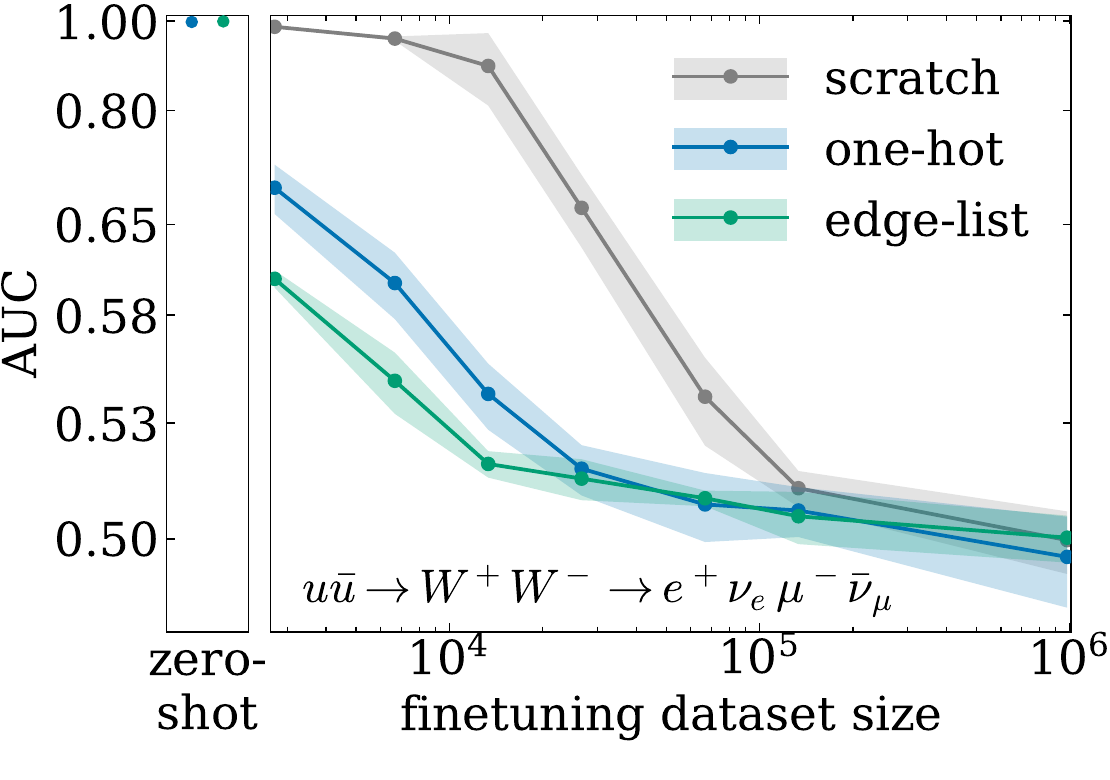}
    \hfill
    \includegraphics[width=0.495\linewidth]{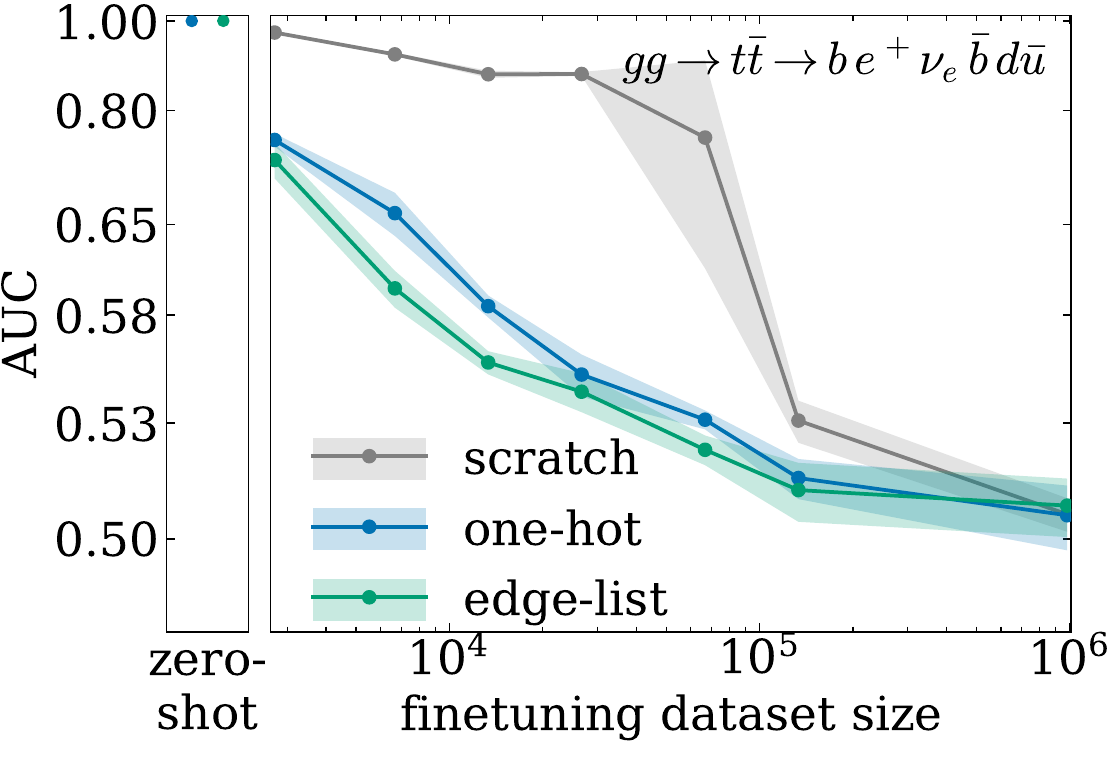}
    \caption{Classifier AUC versus number of finetuning events for high-multiplicity processes, comparing one-hot and edge-list conditioning to training from scratch. The left panel of each plot shows the zero-shot AUC of the pretrained networks. Lines and bands are the mean and standard deviation over five independent runs.}
    \label{fig:finetune_more_particles}
\end{figure}

\begin{figure}[b!]
    \centering
    \includegraphics[width=0.5\linewidth]{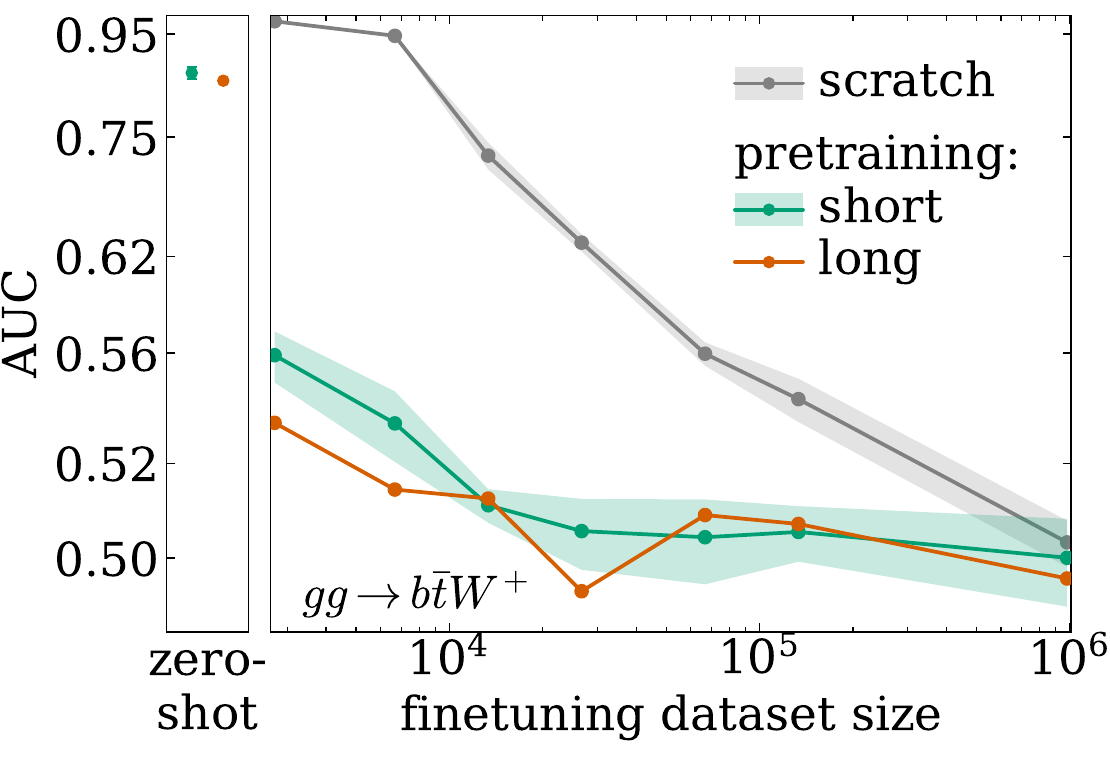}
    \caption{Classifier AUC versus number of finetuning events for $gg \to b\bar t W^+$, comparing finetuning from a standard and an extended edge-list pretraining against training from scratch. The left panel shows the zero-shot AUC. Lines and bands are the mean and standard deviation over five runs; the long pretraining is a single run.}
    \label{fig:finetune_long_auc}
\end{figure}

During pretraining, we use processes with at most three outgoing particles. We now finetune on two decay-chain processes with higher-multiplicity  final states,
\begin{align}
 u\bar u &\to W^+W^- \to e^+\nu_e\,\mu^-\bar\nu_\mu  \notag \\
  gg &\to t\bar t \to b\,e^+\nu_e\,\bar b\,d\bar u 
\label{eq:decay_chain}
\end{align}
Without fine-tuning, pretrained networks cannot generate these final states, and both conditioning schemes lead to an AUC close to one (zero-shot panels of Fig.~\ref{fig:finetune_more_particles}). The edge-list embedding provides no advantage, as it encodes the full final state with its higher multiplicity, but the network has only learned to generate the multiplicities seen during pretraining. After finetuning, the pretraining prior pays off, and the pretrained networks are highly data efficient. They beat the from-scratch baseline using the smallest fine-tuning datasets. As before, the edge-list embedding keeps a small advantage over the one-hot label at low statistics, and the two converge for large datasets. 

In Fig.~\ref{fig:finetune_long_auc}, we ask whether pretraining the backbone longer improves the finetuning. We finetune the hold-out process $gg \to b\bar t W^+$ from the standard edge-list pretraining and from an extended one and show the classifier AUC as a function of the number of finetuning events. Both pretrainings are far more data-efficient than training from scratch, but the short and the long pretraining reach the same AUC across the whole range: a longer pretraining does not improve the classifier AUC.

\begin{figure}[t]
    \includegraphics[width=0.495\linewidth, page=1]{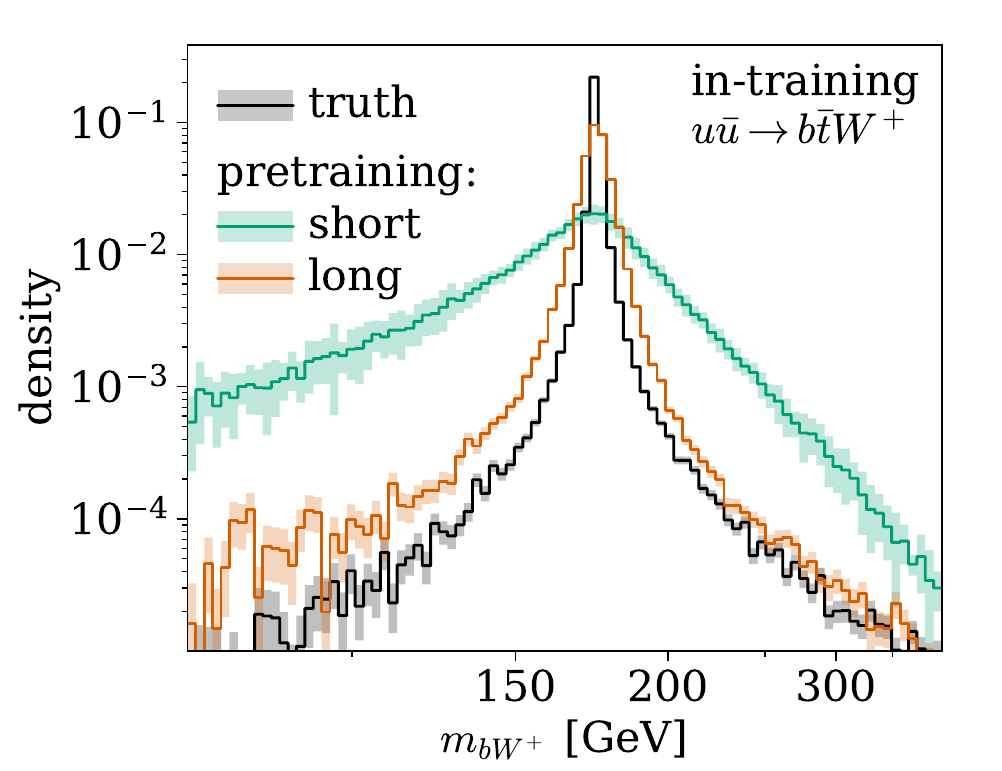}
    \includegraphics[width=0.495\linewidth, page=1]{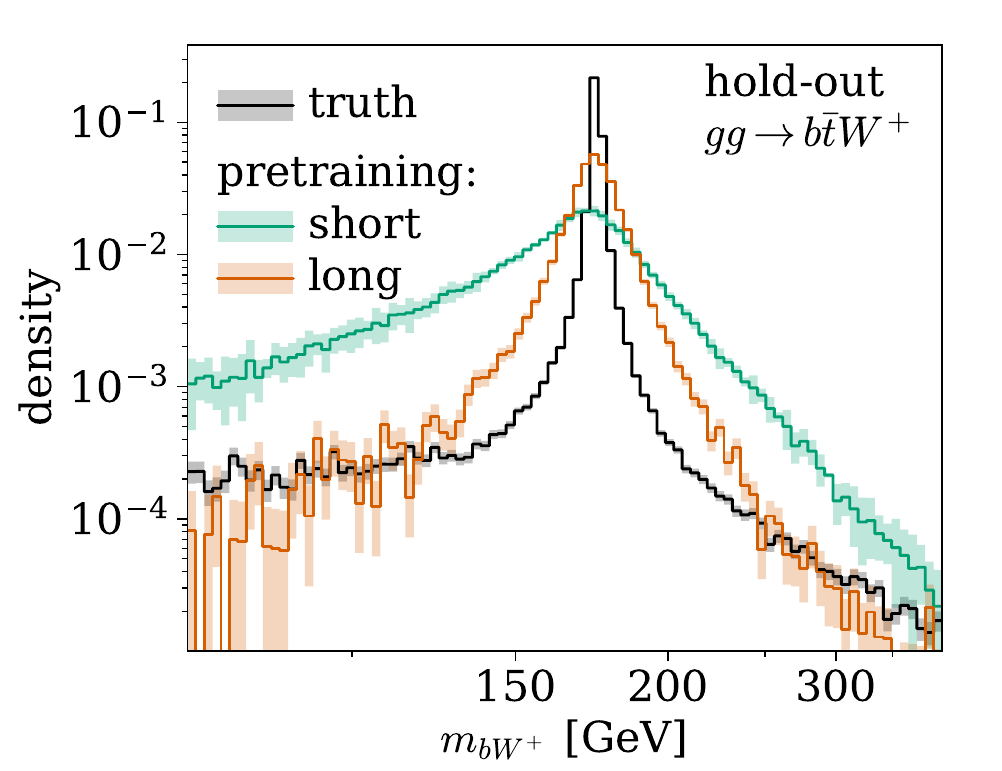}
    \caption{Intermediate $m_{bW^+}$ for in-training $u\bar u \to b\bar t W^+$ (left) and hold-out $gg \to b\bar t W^+$ (right), the latter finetuned on $2048$ events. Green and orange lines denote short and long pretraining, black is the truth. Bands show the run-to-run standard deviation, or Poisson uncertainties for the single long-pretraining run.}
    \label{fig:finetune_long_resonance}
\end{figure}

Finally, we can check the intermediate top resonance on a continuum spanning several hundred GeV, a particularly challenging correlation~\cite{Butter:2019cae}. In the left panel of Fig.~\ref{fig:finetune_long_resonance}, we see that even for an in-training process the standard pretraining never learns the peak, only the extended pretraining captures it. This learned resonance is then transferred across processes. For the hold-out process $gg \to b\bar t W^+$ in the right panel, finetuned on $2048$ events, the long-pretrained network roughly recovers the peak. The full progression from zero-shot through larger finetuning samples is shown in App.~\ref{app:supplementary_results}. This indicates that a descriptively pretrained network can be finetuned for a new process with few events, thanks to the process embedding transferring reusable structure rather than memorizing individual processes.

\section{Outlook}
\label{sec:conclusions}

The choice of representation is key for the generalization properties of a neural network. While the representation of 4-momenta is well understood, the encoding of high-level information correlating different processes remains an open question. We have addressed it employing a multi-modal setup in which a pretrained LLM backbone conditions an autoregressive, conditional flow matching head on high-level information encoded based on a one-hot label, text  strings, Feynman-diagram edge lists, and rendered diagram images. The physics inductive bias from the descriptive embeddings significantly improved the generator performance.

First, we have varied the $Z$ boson mass in Drell-Yan production and conditioned on its numerical value. A one-hot label cannot place the mass peak for a value absent from the training data, while a small mass adapter and a sufficiently capable LLM backbone interpolate to unseen values, and the mass adapter even extrapolates within a certain range.

Second, we have conditioned on process information, using 1491 $2\to 2$ and $2\to 3$ tree-level LHC processes. Descriptive embeddings converged substantially faster than the one-hot baseline and reached much better zero-shot performance on training and hold-out processes. The performance improved with the descriptiveness of the embedding: adding information about the intermediate particles improved over just incoming and outgoing particles. Encoding the Feynman diagrams further improved performance, where the image encoding performed on par with the edge-list encoding of the same information. 

Generating $t \bar{t}H$ events, the embeddings correctly captured qualitative changes between the quark- and gluon-induced channels, of which only the quark-induced channels were used for training. Features like the symmetric rapidity of the gluon-fusion process were transferred, while the generative network struggled in the matrix-element-sensitive threshold region. This shows how descriptive process representations allow transformers to learn high-level correlations across full process catalogs. Combined with finetuning we found a vast efficiency gain over learning each process separately.

\subsection*{Code availability}

The code for this project is published at \url{https://github.com/heidelberg-hepml/LLM4LHC}.

\subsection*{Acknowledgements}

DS is funded by Germany's Excellence Strategy EXC~2181/1 -- 390900948 (the \textsl{Heidelberg STRUCTURES Excellence  Cluster}).  The authors are supported by the Deutsche Forschungsgemeinschaft (DFG, German Research Foundation) under grant 396021762 -- TRR~257 \textsl{Particle Physics Phenomenology after the  Higgs Discovery} and by the KISS (05D23GU4) and SciFM (05D25VH3) consortia funded by BMFTR in the ErUM-Data action plan. The authors acknowledge support by the state of Baden-Württemberg through bwHPC and the German Research Foundation (DFG) through grant no INST 39/963-1 FUGG (bwForCluster NEMO).

\clearpage
\appendix
\section{Technical details} \label{app:technical_details}

\subsection*{Network architecture}

In the following, we describe the CFM head, the pretrained LLM backbones, and the classifier used to evaluate the trained network.

\paragraph{CFM head} \label{app:CFM}

The CFM head, predicting the velocity field $v_\theta(\tilde{k}_i, \tau, k_{[1 : i-1]}, \; \text{process})$ in Eq.~\eqref{eq:cfm_ode}, is an MLP network with residual connections and layer-normalizations. Following the adaptive layer-normalization design of Ref.~\cite{peebles2023scalable}, the two inputs $\tilde{k}_i$ and $(\tau, k_{[1 : i-1]}, \; \text{process})$ enter the network in different ways. The former is passed directly to the MLP, whereas the latter is embedded into a style vector $s \in \rbb^{d_s}$, which modifies the operations of the residual connections and the layer-normalizations.

We construct the style vector as follows. First, the ODE time $\tau$ is mapped to Gaussian Fourier features \cite{Tancik:2020pws}
\begin{align}
    \left( \sin 2\pi \nu_1 \tau , \dots, \sin 2\pi \nu_{d_\tau / 2} \tau, \cos 2\pi \nu_1 \tau, \dots, \cos 2\pi \nu_{d_\tau / 2} \tau  \right)^T \in \rbb^{d_\mathrm{\tau}}\;.
\end{align}
At network initialization, the frequencies $\nu$ are randomly sampled from a Gaussian distribution, $\nu_i \sim \normal(0,\sigma^2)$ with $\sigma=10$, and then held fixed. These features are passed through a small MLP to obtain the time embedding $\tilde{\tau} \in\rbb^{d_\tau}$. The dependence on $(k_{[1 : i-1]}, \; \text{process})$ comes from the transformer output, which we linearly project to $c \in \rbb^{d_c}$.  The style vector is obtained by concatenating the two embeddings, $s = (\tilde{\tau}, c) \in \rbb^{d_s}$ with $d_s = d_\tau + d_c$.

In each MLP block, the style vector is non-linearly mapped to three vectors $\alpha(s)$, $\beta(s)$, and $\gamma(s)$ of the head's hidden dimension $d_h$. The vector $\alpha$ gates the residual connections according to
\begin{equation}
    x \mapsto x + \alpha(s) \odot f_\theta(x)\;,
\end{equation}
where $\odot$ denotes the element-wise multiplication, $x$ the hidden state and $f_\theta$ the nonlinear transformation within the corresponding MLP block. The vectors $\beta(s)$ and $\gamma(s)$ modify the Root-Mean-Square layer-normalization~\cite{zhang2019root} as
\begin{equation}
    x \rightarrow \gamma(s) \odot \frac{x}{\sqrt{\frac{1}{d_h} \sum_{i=1}^{d_h} x_i^2 + \delta}} + \beta(s)\;.
\end{equation}
Here, $\delta > 0$ is a regulator. $\alpha$ and $\beta$ are initialized as zero, whereas $\gamma$ is initialized to unity.

\paragraph{Large Language Models} \label{app:llms}

We use the instruction-tuned LLMs \href{https://huggingface.co/Qwen/Qwen2.5-0.5B-Instruct}{Qwen2.5-0.5B}~\cite{yang2024qwen2} and \href{https://huggingface.co/Qwen/Qwen3-VL-2B-Instruct}{Qwen3-VL-2B}~\cite{qwen3vl2025}. Both are decoder-only transformers with causal self-attention that share essentially the same architecture: a stack of transformer blocks, each with a grouped-query self-attention layer using rotary position embeddings and a SwiGLU feed-forward layer. Both also use the same tokenizer and vocabulary.

The smaller Qwen2.5-0.5B supports only the textual modality, while the larger Qwen3-VL-2B model additionally supports image and video inputs. Qwen3 introduces several architectural improvements over Qwen2.5, the main one being the multimodal rotary position embeddings. A vision encoder maps the visual inputs to a sequence of visual tokens. The encoder consists of a single convolutional layer, a vision transformer, and a downsampling layer. The convolutional layer maps each $2\times16\times16$ (time $\times$ height $\times$ width) patch of the three-channel input to a $1024$-dimensional vector, which serve as the input to the vision transformer. The output patches of the vision transformer are then merged in $2\times2$ groups and projected into the input space of the transformer backbone. Images are effectively treated as single-frame videos, with the frame repeated to fill the temporal patch dimension of the convolutional layer.

The network hyperparameters of both LLMs are listed in Tab.~\ref{tab:llm_hyperparams}. For architectural details, we refer the reader to the respective references. When loading the pretrained backbone weights, we reset the final layer-normalization weights to unity instead of using their pretrained values, following Ref.~\cite{Heneka:2025fpe}.

\begin{table}[b!]
    \centering
    \begin{small}
    \begin{tabular}[t]{l |c c}
    \toprule
    Backbone transformer & Qwen2.5-0.5B & Qwen3-VL-2B  \\
    \midrule
    Parameters              & $0.5\,$B   & $2\,$B    \\
    Hidden dimension        & $896$      & $2048$    \\
    Transformer blocks      & $24$       & $28$      \\
    Attention heads         & $14$       & $16$      \\
    Key--value heads        & $2$        & $8$       \\
    Attn. head dimension    & $64$       & $128$     \\
    Feed-forward dimension  & $4864$     & $6144$    \\
    Vocabulary size         & $151\,936$ & $151\,936$ \\
    RoPE base               & $10^{6}$   & $5\times10^{6}$ \\
    Context length          & $32\,768$  & $262\,144$ \\
    \bottomrule
    \end{tabular}
    \qquad
    \begin{tabular}[t]{l |c}
    \toprule
    Vision transformer & Qwen3-VL-2B \\
    \midrule
    Parameters              & \\
    Hidden dimension        & $1024$ \\
    Transformer blocks      & $24$   \\
    Attention heads         & $16$   \\
    Attn. head dimension    & $64$   \\
    Feed-forward dimension  & $4096$ \\
    Output dimension        & $2048$ \\
    \bottomrule
    \end{tabular}
    \end{small}
    \caption{Network hyperparameters of the LLM backbones Qwen2.5-0.5B and
    Qwen3-VL-2B (left) and of the Qwen3-VL vision transformer (right).}
    \label{tab:llm_hyperparams}
\end{table}

\paragraph{Classifier}

For each AUC evaluation, we train a gradient-boosted decision tree classifier with $200$ boosting iterations, a learning rate of $0.1$, and early stopping with a patience of $10$. Its inputs are the final-state momenta in the same coordinates as for the momentum embedding, but with the fixed on-shell masses removed. The two classes, truth and generated samples, are balanced by weighting each event inversely to its class size. This way, an AUC of $0.5$ implies that the classifier cannot distinguish the true and generated samples from each other.

\subsection*{Computational optimizations}

To reduce the compute, we apply the following optimizations.

Each input sequence consists of the process-encoding prefix tokens followed by the particle momenta. Since the prefix tokens pass through the frozen LLM weights, we cache their key-value activations and compute them ahead of time. To make this caching efficient, all events in a single training batch are from a single process only.

Following hydragen~\cite{juravsky2024hydragen}, in the forward pass the momenta attend to the cached prefix and to themselves separately, recombined by an online softmax. In the backward pass, the frozen prefix keys and values receive no gradient, so only the gradient with respect to the queries is needed. We compute it with a custom Triton kernel that returns the query gradient alone, skipping the prefix key and value gradients.

We also make use of the CFM head being much smaller than the backbone. For each predicted momentum we draw multiple noise and ODE time samples used for the CFM loss in Eq.~\eqref{eq:cfm_ode}. As the autoregressive conditioning is shared across those samples, the backbone is evaluated only once per event. Before backpropagating through the backbone, we aggregate the gradients from all samples of the same event.

Finally, the backbone runs in its native bfloat16 numerical precision, whereas the CFM head and all remaining parameters run in float32 to preserve numerical precision for the continuous physical quantities.

\subsection*{Hyperparameter}

\subsubsection*{Parameter conditioning}

Table~\ref{tab:dy_hyperparams} lists the network and training hyperparameter shared by all Drell-Yan parameter-conditioning runs. The one-hot encoding and mass-adapter add a particle-type embedding to the momenta, which the text scheme omits. For the one-hot encoding, the embedding finetuning of the hold-out mass is performed for $2$ epochs at batch size $128$. The mass adapter is a two-layer MLP ($1 \to 128 \to 896$).

\begin{table}[b!]
    \centering
    \begin{small}
    \begin{minipage}[t]{0.4\linewidth}
    \centering
        \begin{tabular}[t]{l |c}
        \toprule
        \multicolumn{2}{l}{CFM head} \\
        \midrule
        Parameters                & $3.3\,$M\\  
        Hidden dimension          & $256$ \\
        Blocks                    & $4$ \\
        Time-embedding dim.       & $64$ \\
        Conditioning dim.         & $256$\\
        Dropout                   & $0.1$ \\
        \bottomrule
        \end{tabular}

        \vspace{12pt}

        \begin{tabular}[t]{l |c}
        \toprule
        \multicolumn{2}{l}{Sampling} \\
        \midrule
        ODE solver                & dopri5 \\
        rtol                      & $10^{-5}$ \\
        atol                      & $10^{-7}$ \\
        Samples per mass          & $100$k \\
        \bottomrule
        \end{tabular}
    \end{minipage}
    \qquad
    \begin{minipage}[t]{0.4\linewidth}
    \centering
        \begin{tabular}[t]{l |c}
        \toprule
        \multicolumn{2}{l}{Training} \\
        \midrule
        Optimizer                   & AdamW \\
        Learning rate               & $10^{-4}$ \\
        Weight decay                & $0.005$ \\
        Gradient clipping           & $1.0$ \\
        Batch size                  & $2048$ \\
        Gradient acc.               & $2$ \\
        Epochs                      & $400$ \\
        Noise samples per momentum  & $16$ \\
        \bottomrule
        \end{tabular}

        \vspace{12pt}

        \begin{tabular}[t]{l |c}
        \toprule
        \multicolumn{2}{l}{Data} \\
        \midrule
        Events per mass           & $100$k \\
        Train/val/test            & $0.7 / 0.1 / 0.2$ \\
        \bottomrule
        \end{tabular}
    \end{minipage}
    \end{small}
    \caption{Network, training, and sampling hyperparameter shared by the
    Drell-Yan parameter-conditioning runs.}
    \label{tab:dy_hyperparams}
\end{table}

\subsubsection*{Process conditioning}

Table~\ref{tab:proc_hyperparams} lists the network and training hyperparameter
shared by all process-conditioning runs.

\begin{table}[t]
    \centering
    \begin{small}
    \begin{minipage}[t]{0.4\linewidth}
    \centering
        \begin{tabular}[t]{l |c}
        \toprule
        \multicolumn{2}{l}{CFM head} \\
        \midrule
        Parameters                & \makecell{$3.3\,$M (Qwen2)\\$6.2\,$M (Qwen3)}\\
        Hidden dimension          & $128$ \\
        Blocks                    & $6$ \\
        Time-embedding dim.       & $64$ \\
        Conditioning dim.         & \makecell{$896$ (Qwen2)\\$2048$ (Qwen3)} \\
        Dropout                   & $0.1$ \\
        \bottomrule
        \end{tabular}

        \vspace{12pt}

        \begin{tabular}[t]{l |c}
        \toprule
        \multicolumn{2}{l}{Sampling} \\
        \midrule
        ODE solver                & dopri5 \\
        rtol                      & $10^{-5}$ \\
        atol                      & $10^{-7}$ \\
        Samples per process       & $20$k\\
        \bottomrule
        \end{tabular}
    \end{minipage}
    \qquad
    \begin{minipage}[t]{0.4\linewidth}
    \centering
        \begin{tabular}[t]{l |c}
        \toprule
        \multicolumn{2}{l}{Training} \\
        \midrule
        Optimizer                   & AdamW \\
        Learning rate               & $5 \times 10^{-5}$ \\
        Weight decay                & $0.005$ \\
        Gradient clipping           & $1.0$ \\
        Batch size                  & $6144$ \\
        Gradient acc.               & $4$ \\
        Epochs                      & $20$ \\
        Noise samples per momentum  & $10$ \\
        \bottomrule
        \end{tabular}

        \vspace{12pt}

        \begin{tabular}[t]{l |c}
        \toprule
        \multicolumn{2}{l}{Data} \\
        \midrule
        Events per process        & $100$k \\
        Train/val/test            & $0.6 / 0.2 / 0.2$ \\
        \bottomrule
        \end{tabular}
    \end{minipage}
    \end{small}
    \caption{Network, training, and sampling hyperparameter shared by the
    process-conditioning runs. The conditioning dimension of the CFM head equals the backbone hidden dimension.}
    \label{tab:proc_hyperparams}
\end{table}

The one-hot encoding adds a particle-type embedding to the momenta, which the text and image schemes omit. For the hold-out processes, its embeddings are finetuned for $20$ epochs.

\paragraph{Long pretraining}

The long pretraining runs train for $120$ epochs. For the first $60$ epochs each process contains 60k training events. For the second half, we extend the per-process training data size to roughly $10^6$ events, and randomly draw 60k training events in each epoch, such that the optimization updates within an epoch stays the same. For the final $20$ epochs the learning rate is decayed with a cosine annealing scheduler and the noise samples per momentum are increased from $10$ to $32$.

\paragraph{Non-LLM graph embedding}

The graph-encoder is described in App.~\ref{app:Feynman_diagram_input}. The runs use the same CFM head and training settings as in Tab.~\ref{tab:proc_hyperparams} and train for $80$ epochs. The backbone is a small transformer trained from scratch with hidden dimension $256$, $6$ blocks, $8$ attention heads, and a SwiGLU feed-forward dimension of $682$, about $5$M parameters. Like the one-hot encoding, they carry the particle-type embedding.

\paragraph{Finetuning}

For the finetuning runs, we keep the test set fixed at $20$k events and vary the size of the training and validation sets. We train for $20$ epochs with $16$ noise samples per momentum and use a cosine annealing learning rate scheduler. The batch size is $1024$. The network trained from scratch uses the same architecture and pretrained LLM backbone weights, with a single one-hot encoding and the particle-type embedding.

\section{Feynman diagram input}
\label{app:Feynman_diagram_input}

\begin{figure}[b!]
    \centering
    \includegraphics[width=0.5\linewidth]{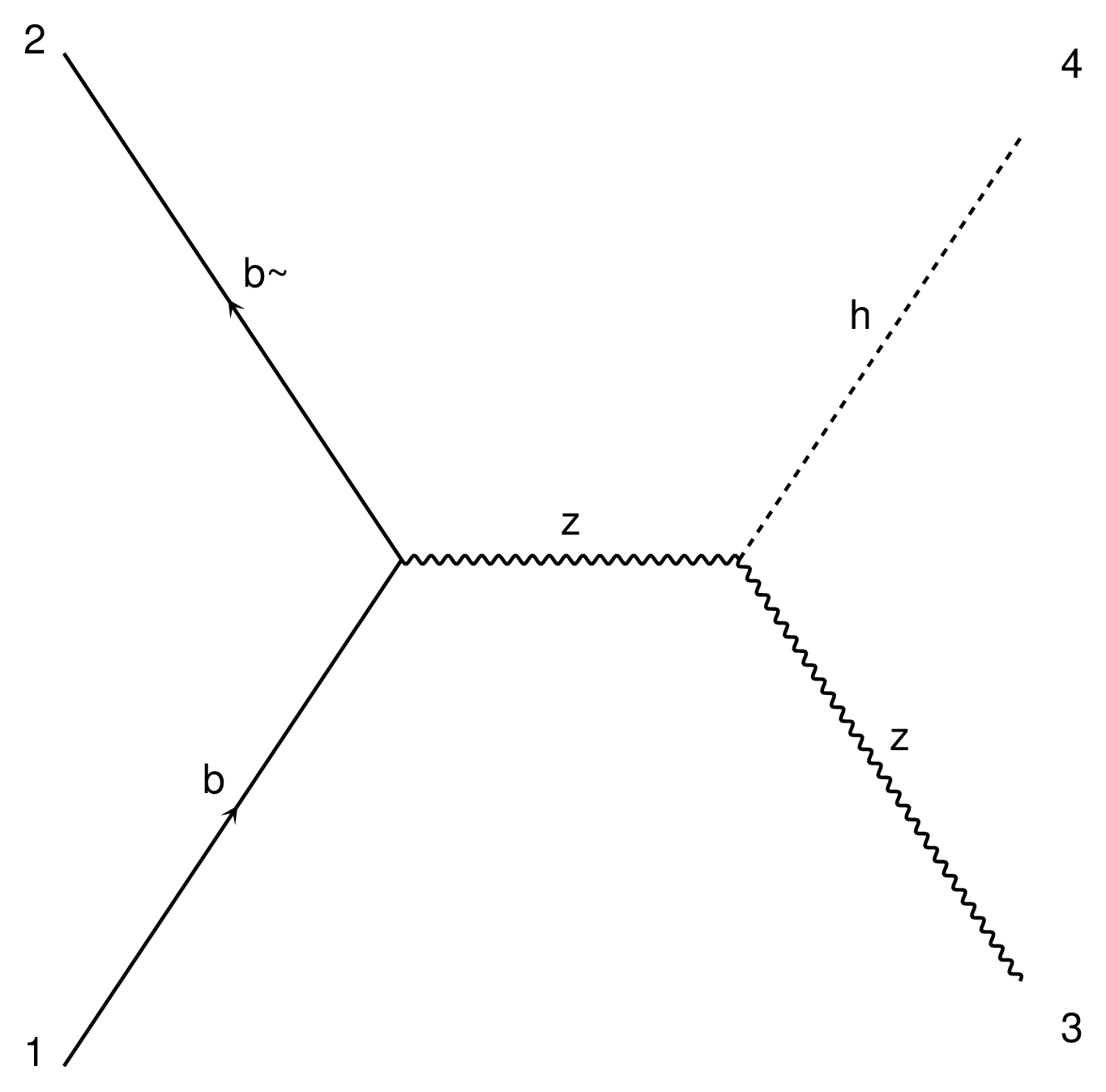}
    \caption{Exemplary Feynman diagram image used as input.}
    \label{fig:diagram_example}
\end{figure}

We show an exemplary Feynman diagram image for the $b\bar b\to Z H$ process in Fig.~\ref{fig:diagram_example}. This image is taken directly from \madgraph and passed as input to the Qwen3-VL vision encoder.

The associated edge list string reads ``b(1)-b\scalebox{0.6}[1]{\textasciitilde}(2)-V0\textasciitilde z\textasciitilde z(3)-h(4)-V1''. The numbers in the brackets are used to denote the external particles. The dashes are used to denote connections to internal vertices, which are labelled as ``V0'', ``V1'', $\ldots$. The ``\textasciitilde z \textasciitilde'' is used to denote the internal $Z$ boson propagator connecting the V0 and V1 vertices.

\subsection*{Graph encoder}

The graph modality of Sec.~\ref{sec:setup} encodes each Feynman diagram with a small graph transformer that acts on one diagram at a time. The nodes are the external particles --- split into incoming and outgoing --- together with the internal vertices, and the edges are the external legs and internal propagators. Each node carries its type and a Laplacian positional encoding~\cite{dwivedi2020generalization}. Each edge carries the identity of the particle it represents.

The node-type embeddings are passed through several layers of multi-head self-attention over the nodes, in which the particle on each edge enters through learned terms~\cite{shaw2018selfattention}: a scalar $\beta_{\ell_{ij}}$ added to the attention logit, and vectors $r^K_{\ell_{ij}}, r^V_{\ell_{ij}}$ added to the key and value of the connected nodes. In this way, the identity of the propagating particle is aggregated into the node states. Concretely, in each head the node states $h_i$ are linearly projected to queries, keys, and values $q_i, k_i, v_i \in \rbb^{d_h}$, and the attention score, weight, and node update read
\begin{align}
    e_{ij} = \frac{q_i^\top \bigl( k_j + r^K_{\ell_{ij}} \bigr)}{\sqrt{d_h}} + \beta_{\ell_{ij}} \;, \qquad
    \alpha_{ij} = \softmax_j \bigl( e_{ij} \bigr) \;, \qquad
    \tilde{h}_i = \sum_j \alpha_{ij} \bigl( v_j + r^V_{\ell_{ij}} \bigr) \; .
\end{align}
Here $\ell_{ij}$ labels the particle on the edge between nodes $i$ and $j$, and the edge terms $\beta_{\ell_{ij}} \in \rbb$ and $r^K_{\ell_{ij}}, r^V_{\ell_{ij}} \in \rbb^{d_h}$ are learned per particle type and shared across layers. They are masked by the graph adjacency and vanish for node pairs not connected by a leg or propagator, where the update reduces to standard dot-product attention. A learnable readout token collects each diagram into a single latent vector $E_\theta(\text{diagram}) \in \rbb^d$. 

The graph transformer has $3$ layers, $4$ heads, a hidden dimension of $256$, and a Laplacian positional encoding of dimension $8$.

\section{Supplementary results}
\label{app:supplementary_results}

\subsection*{Parameter conditioning}

\begin{figure}[b!]
    \includegraphics[width=0.495\linewidth, page=2]{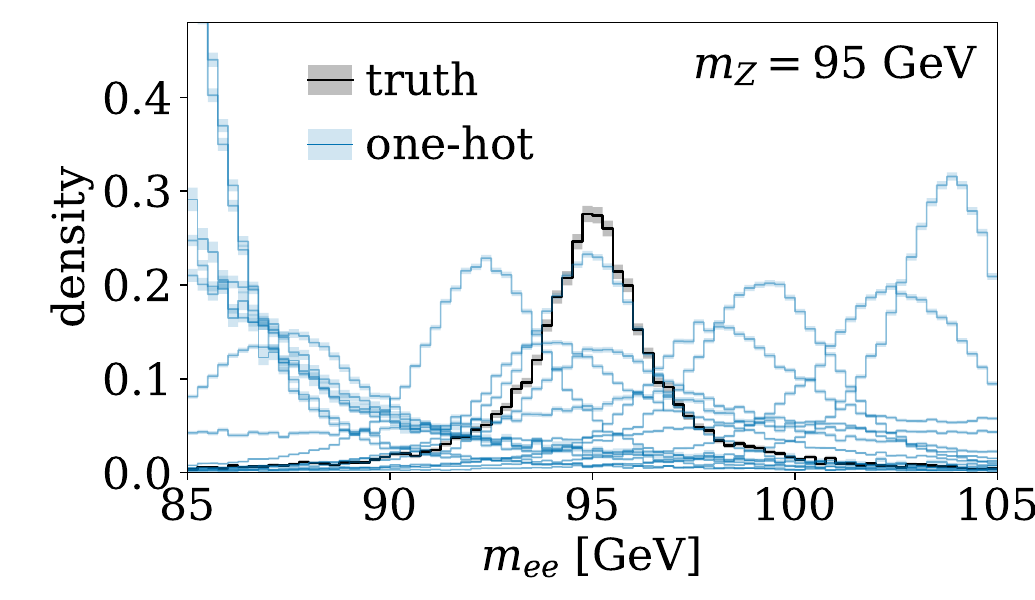}
    \includegraphics[width=0.495\linewidth, page=2]{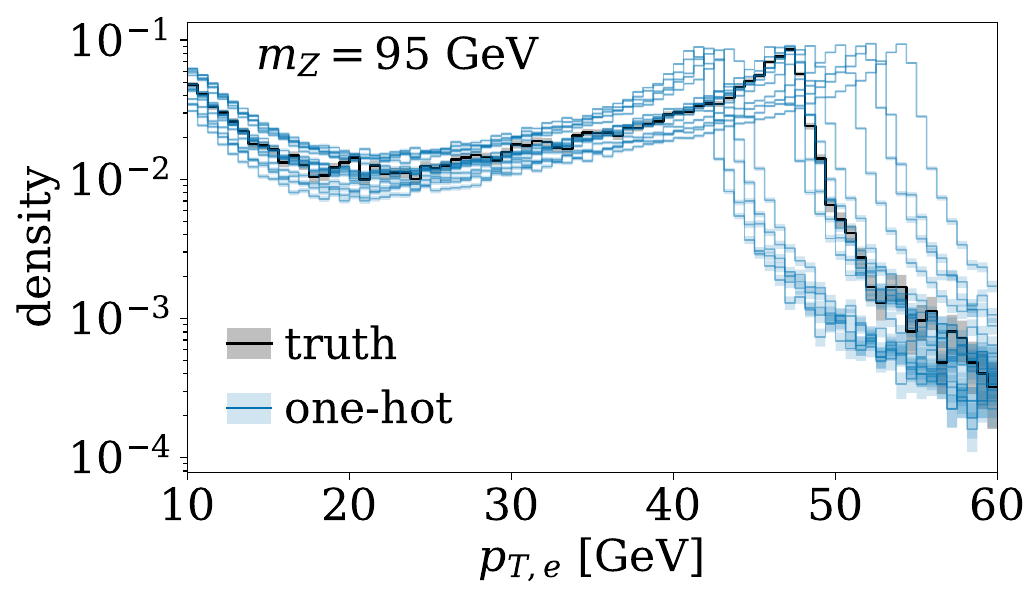}
    \caption{Left: Drell-Yan invariant mass distribution. Right: same for the positron transverse momentum. The truth distribution is shown for $m_Z = 95\,\gev$, and the predicted distributions are obtained by randomly initialized an embedding vector.}
    \label{fig:param_random_embd}
\end{figure}

In Fig.~\ref{fig:param_random_embd}, we show the resulting distributions of the one-hot conditioning, where the embedding vector for the $95\,\gev$ $Z$-mass is randomly initialized. For each of the five one-hot runs of Sec.~\ref{sec:continous_conditioning}, we sample the embedding vector twice. As the random embedding does not encode the hold-out mass, the predicted invariant mass and transverse momentum distributions scatter widely.

\begin{figure}[t]
    \includegraphics[width=0.495\linewidth, page=3]{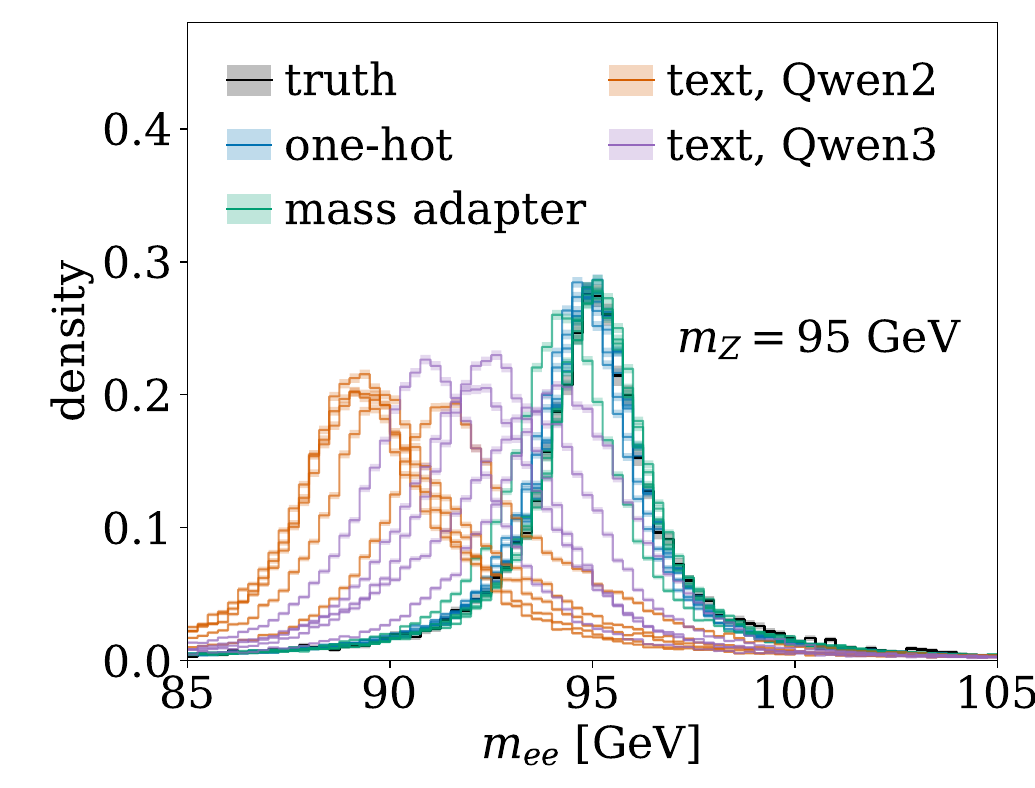}
    \hfill
    \includegraphics[width=0.495\linewidth, page=3]{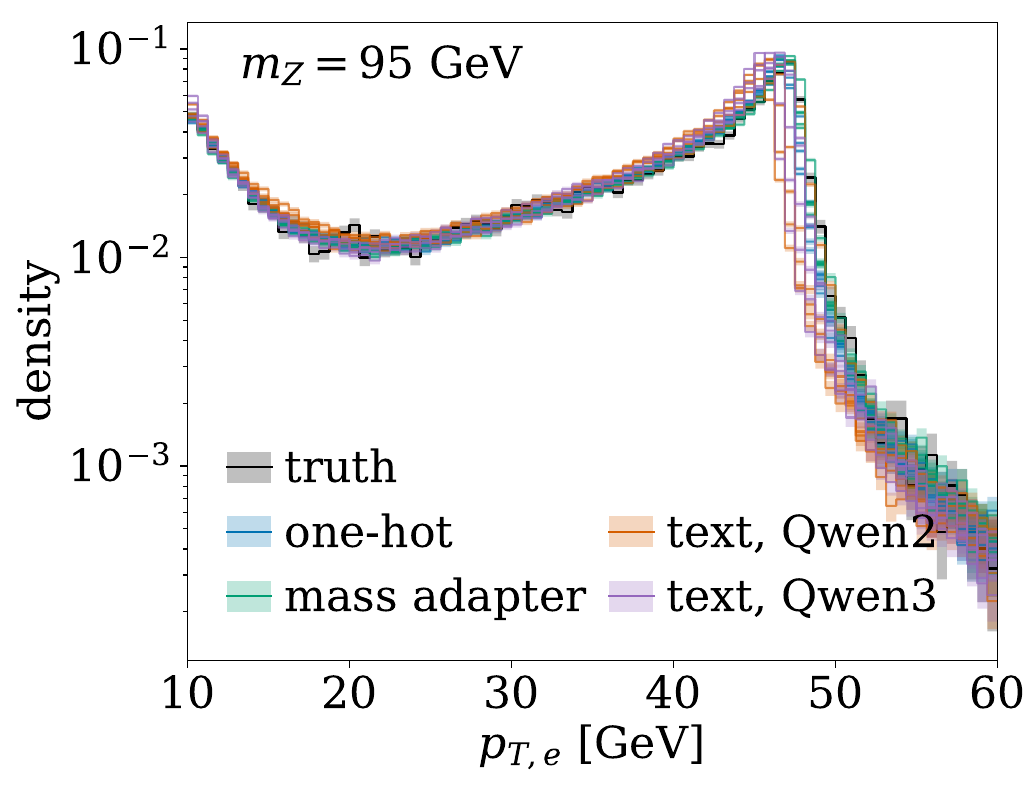}
    \caption{Drell-Yan invariant mass (left) and positron transverse momentum (right) at the interpolated mass $m_Z = 95\,\gev$, showing the five independent runs individually. This is the per-run version of the lower panels of Fig.~\ref{fig:DY}, where the same results are averaged into bands.}
    \label{fig:param_runs}
\end{figure}

In Fig.~\ref{fig:param_runs}, we complement the lower panels of Fig.~\ref{fig:DY} by showing the five independent runs individually rather than averaged into bands. The finetuned one-hot embedding and the mass adapter consistently recover the $Z$ peak across runs, whereas the text conditioning remains shifted towards lower masses and exhibits a sizeable run-to-run spread.

\subsection*{AUC overviews for $2\to2$ and $2\to3$ processes}

\begin{figure}[t]
    \includegraphics[width=0.99\linewidth]{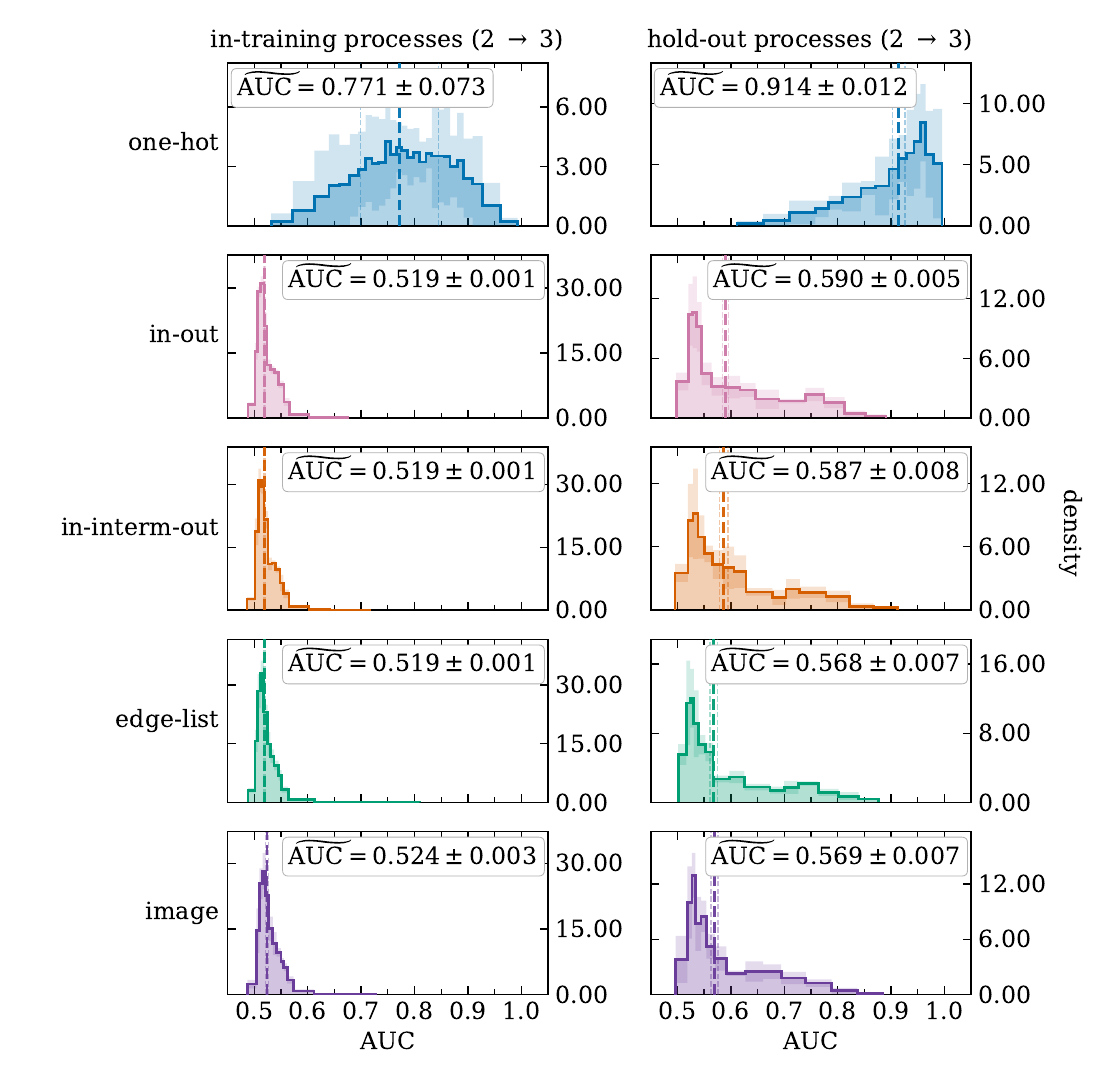}
    \caption{Comparison of the AUCs of different conditioning schemes for the $2\to3$ processes only, with $1116$ processes used for training and $121$ held out. The dashed line indicates the median AUC, with uncertainties from averaging over five independent runs.}
    \label{fig:proc_encoding_comparison_2to3}
\end{figure}

\begin{figure}[t]
    \includegraphics[width=0.99\linewidth]{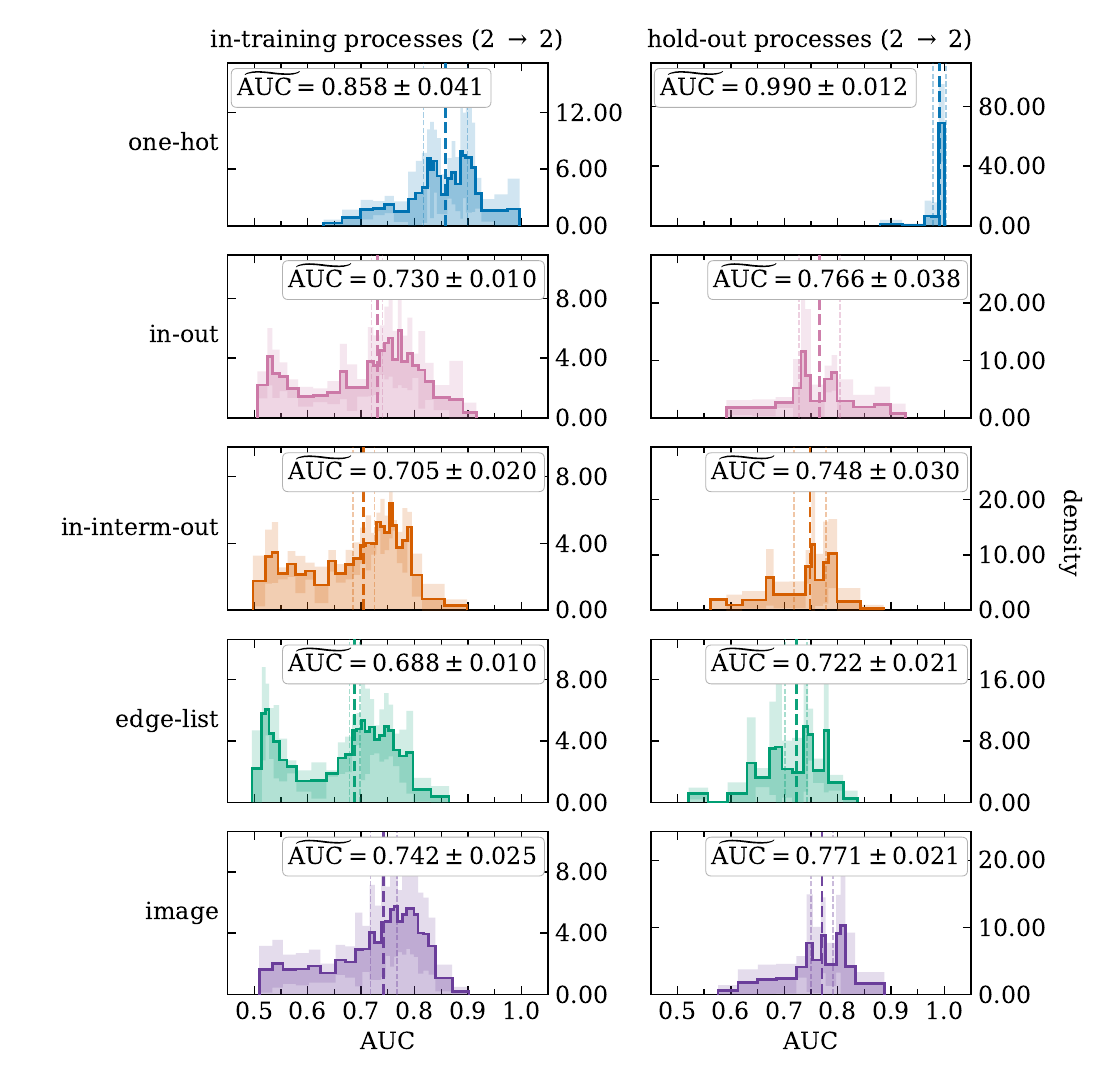}
    \caption{Comparison of the AUCs of different conditioning schemes for the $2\to2$ processes only, with $1116$ processes used for training and $121$ held out. The dashed line indicates the median AUC, with uncertainties from averaging over five independent runs.}
    \label{fig:proc_encoding_comparison_2to2}
\end{figure}

In Figs.~\ref{fig:proc_encoding_comparison_2to3} and \ref{fig:proc_encoding_comparison_2to2}, we restrict the AUC histograms of Fig.~\ref{fig:proc_encoding_comparison} to the $2\to3$ and $2\to2$ processes, respectively. The AUC values are significantly higher for the $2\to2$ processes. The main reason is that the training dataset contains more $2\to3$ than $2\to2$ processes. Moreover, a $2\to2$ final state has an intrinsic dimension of only two, much smaller than the dimension of the space on which the CFM head operates. Such low-dimensional distributions embedded in a higher-dimensional space are notoriously hard for a CFM to learn. The $2\to3$ processes have an intrinsic dimension of five, which is easier to model.

\begin{figure}[htp!]
    \centering
    \includegraphics[width=0.99\linewidth]{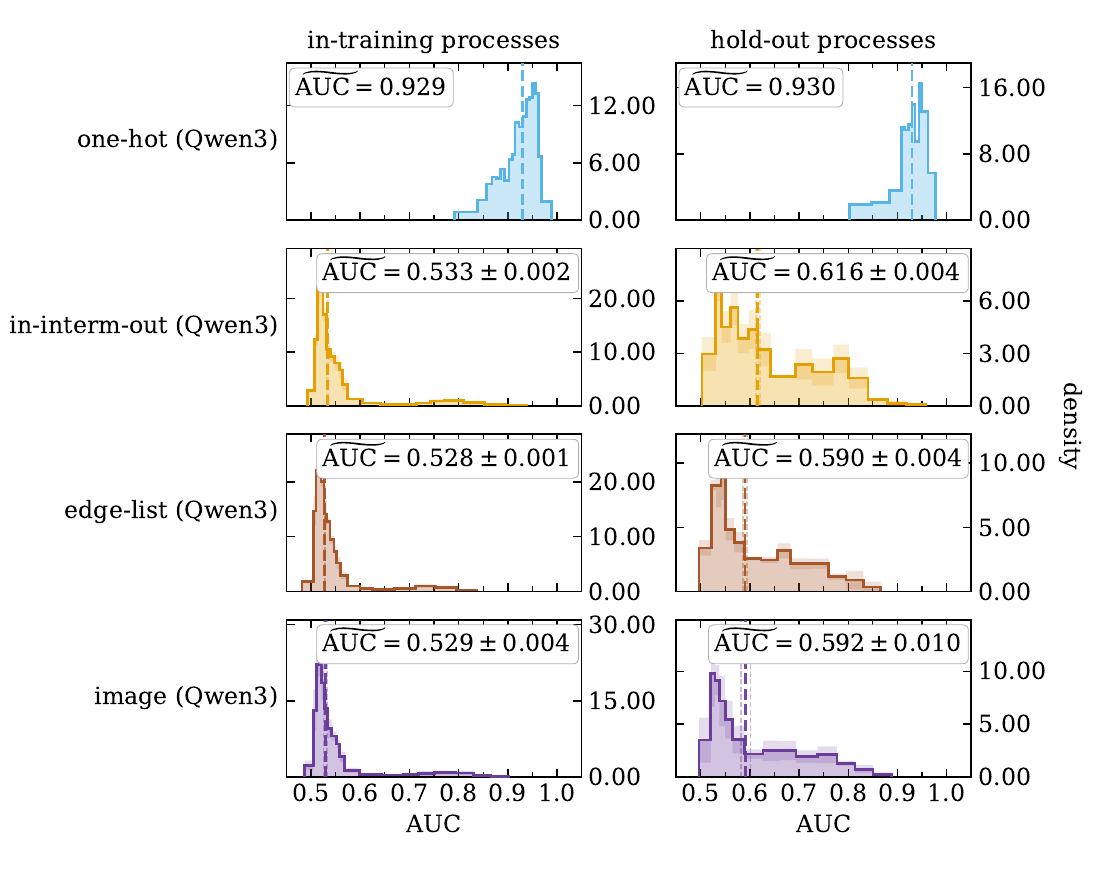}
    \caption{AUCs for the different input conditioning schemes using the Qwen3 backbone in analogy to Fig.~\ref{fig:proc_encoding_comparison}. The dashed line indicates the median AUC, with uncertainties from averaging over independent runs.}
    \label{fig:proc_encoding_comparison_q3}
\end{figure}

Finally, Fig.~\ref{fig:proc_encoding_comparison_q3} repeats the comparison of Fig.~\ref{fig:proc_encoding_comparison} for the Qwen3 backbone with the only exception that we train only a single one-hot run. The ordering of the conditioning schemes is unchanged: the one-hot label fails, the text-based descriptions perform well, and the more descriptive Feynman-diagram encodings further improve the hold-out performance. The image embedding, which requires the Qwen3 backbone, performs on par with the edge-list encoding of the same diagrams.

\subsection*{AUCs of individual processes}

To complement the aggregated histograms of Fig.~\ref{fig:proc_encoding_comparison}, we show the classifier AUC of every individual process for the different input representations. Fig.~\ref{fig:process_comparison} collects the training processes across its four panels, and the following figure shows the hold-out processes. In each panel, blue circles denote the one-hot label, orange squares the in-interm-out string, green triangles the edge-list, and pink diamonds the image representation. The more descriptive embeddings tend to lie below the one-hot baseline, both for the training and the hold-out processes, consistent with the medians of Fig.~\ref{fig:proc_encoding_comparison}.

\begin{figure}[htp!]
    \centering
    \includegraphics[width=0.49\textwidth, page=1]{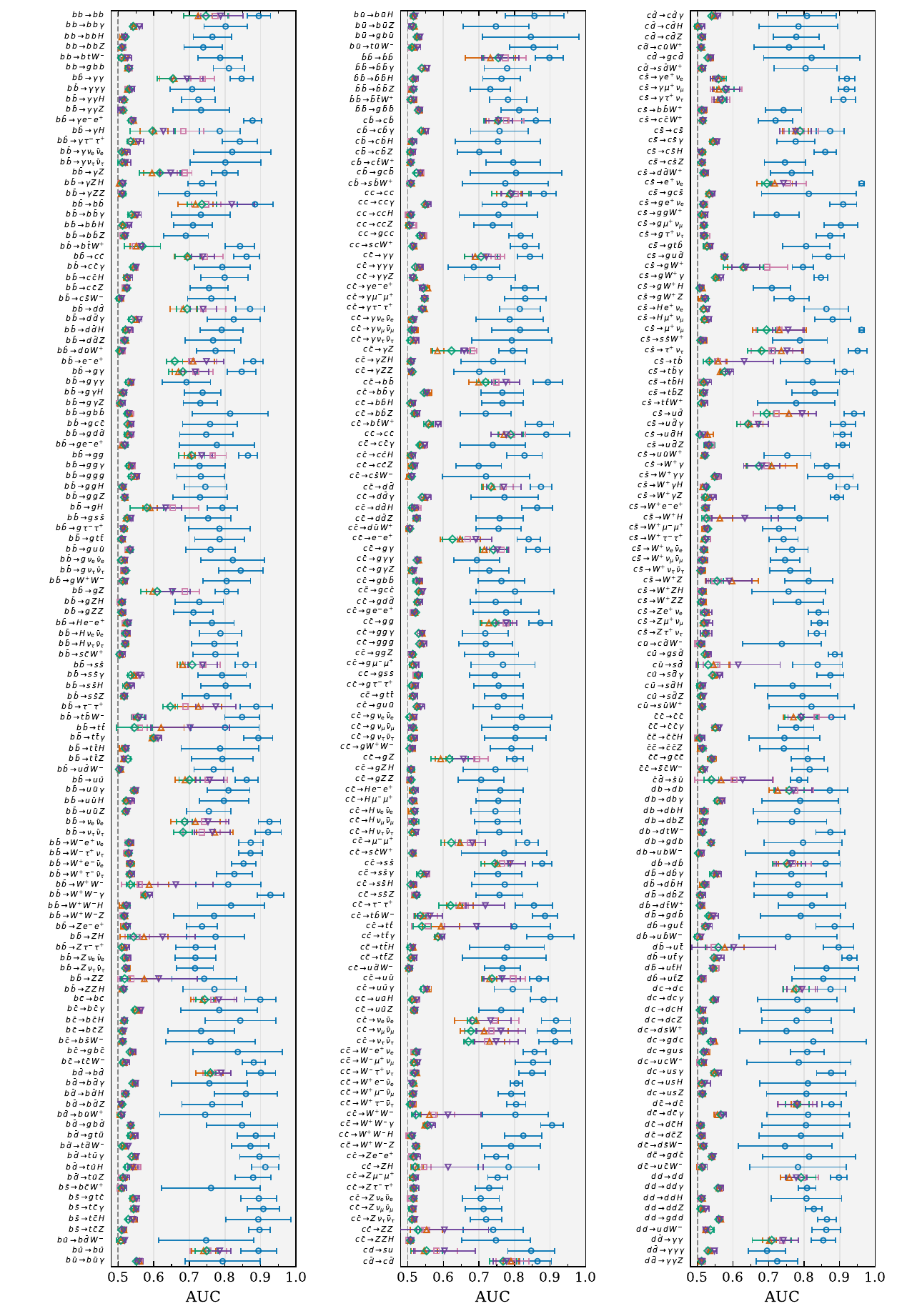}
    \includegraphics[width=0.49\textwidth, page=2]{figs/processes/ManyProcesses/auc_process_comparison.pdf}\\
    \includegraphics[width=0.49\textwidth, page=3]{figs/processes/ManyProcesses/auc_process_comparison.pdf}
    \includegraphics[width=0.49\textwidth, page=4]{figs/processes/ManyProcesses/auc_process_comparison.pdf}
    \caption{Overview of individual train process AUCs for the different input representations. Blue circles denote one-hot, pink squares in-out, orange upward-pointing triangles in-interm-out, green diamonds edge-list, and purple downward-pointing triangles image representation.}
    \label{fig:process_comparison}
\end{figure}

\begin{figure}[htp!]
    \centering
    \includegraphics[width=1.0\textwidth, page=5]{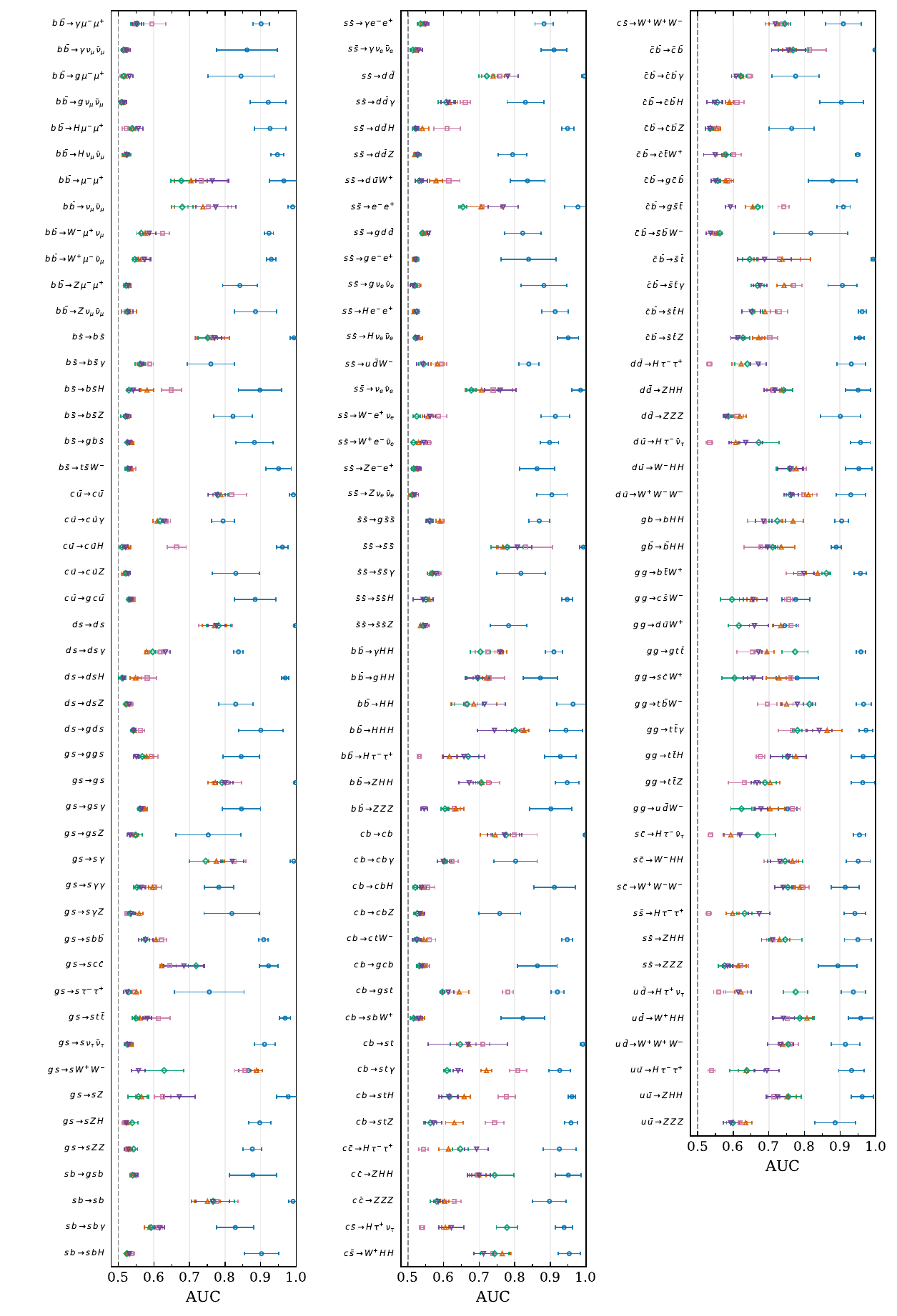}
    \caption{Overview of individual hold-out process AUCs for the different input representations. Blue circles denote one-hot, pink squares in-out, orange upward-pointing triangles in-interm-out, green diamonds edge-list, and purple downward-pointing triangles image representation.}
\end{figure}

\subsection*{Long-training loss curves}

\begin{figure}[htp!]
    \centering
    \includegraphics[width=.5\textwidth]{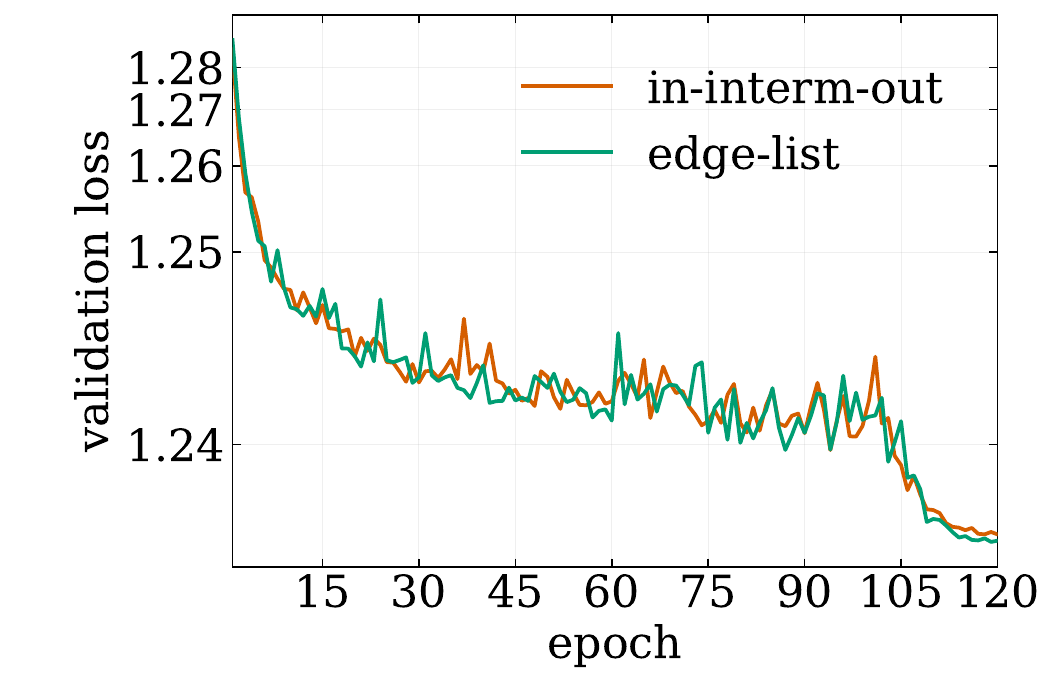}
    \caption{Validation loss over 120 epochs for the in-interm-out and edge-list conditioning, corresponding to the long-training runs of Fig.~\ref{fig:auc_long}.}
    \label{fig:val_loss_long}
\end{figure}

For completeness, Fig.~\ref{fig:val_loss_long} shows the validation loss of the long-training runs of Fig.~\ref{fig:auc_long}, for the in-interm-out and edge-list conditioning over the full 120 epochs. Both schemes track each other closely throughout training and decrease steadily, with a marked drop in the final epochs due to the final cosine annealing scheduling.

\subsection*{Top resonance across finetuning dataset sizes}

\begin{figure}[htp!]
    \centering
    \includegraphics[width=\linewidth, page=1]{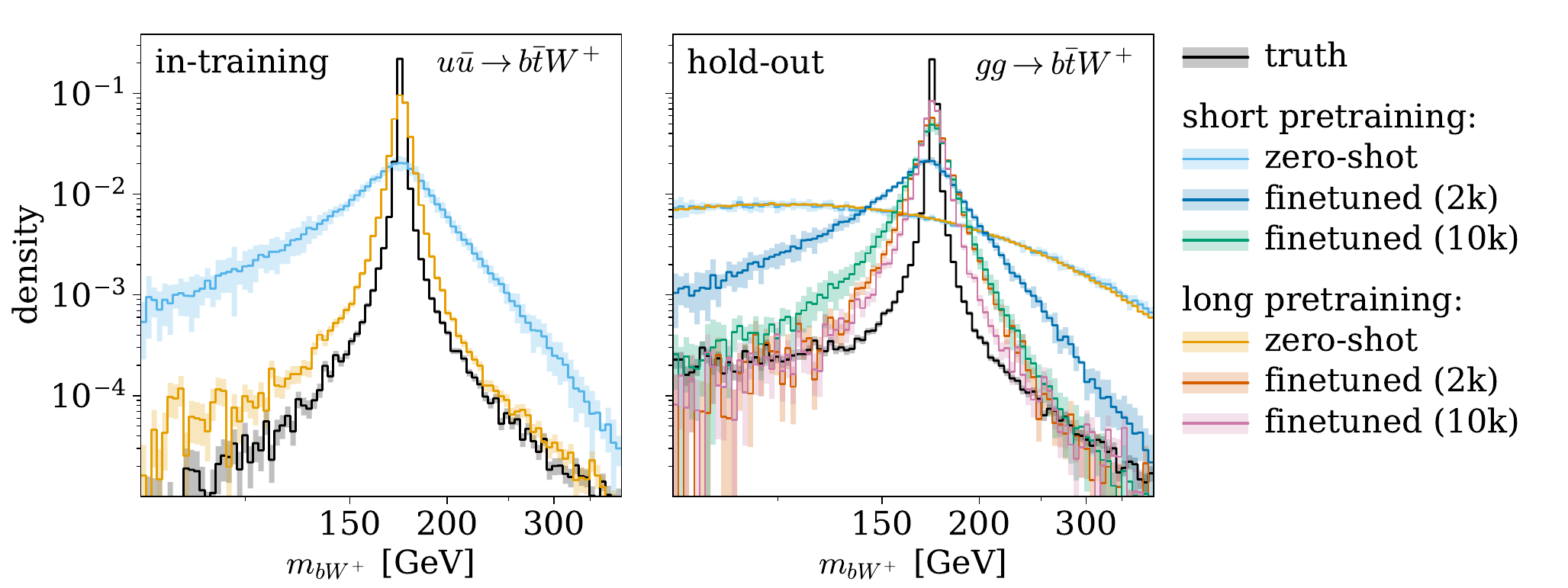}
    \caption{Invariant mass $m_{bW^+}$ reconstructing the intermediate top resonance, for the in-training $u\bar u \to b\bar t W^+$ (left) and the hold-out $gg \to b\bar t W^+$ (right), the latter shown pretrained (zero-shot) and finetuned on $2048$ and $10000$ events. Line colour denotes the pretraining length (short/long) and the finetuning dataset size, as in the legend, black denotes the truth. Bands show the run-to-run standard deviation, or Poisson uncertainties for the single long-pretraining run.}
    \label{fig:finetune_long_resonance_app}
\end{figure}

Fig.~\ref{fig:finetune_long_resonance_app} extends the right panel of Fig.~\ref{fig:finetune_long_resonance} across finetuning dataset sizes. In case of zero-shooting, the $(b,W^+)$ peak is absent for both pretrainings. Finetuning progressively recovers it with the long pretraining performing better than the short one in particular for the smallest dataset size.


\clearpage
\bibliography{tilman, references}

\end{document}